\documentclass[amsmath,aps,prl,twocolumn,longbibliography,superscriptaddress,floatfix]{revtex4-2}
\usepackage[usenames,dvipsnames]{xcolor}
\definecolor{virdark}{HTML}{440154}
\definecolor{virmiddle}{HTML}{218F8D}
\definecolor{viryellow}{HTML}{FDE725}
\definecolor{mygray}{HTML}{AAAAAA}
\definecolor{viryellowdark}{HTML}{807312}

\usepackage{graphicx}
\usepackage{amsmath}
\usepackage{amssymb}
\usepackage{bm}
\usepackage{enumerate}
\usepackage{xfrac}
\usepackage[caption=false]{subfig}
\usepackage[normalem]{ulem}
\usepackage[usenames,dvipsnames]{xcolor}
\usepackage{txfonts}
\usepackage[mathscr]{euscript}
\usepackage[pdftex,
			pdflang={en-US},
            colorlinks=true,
            citecolor=virmiddle,
            linkcolor=virmiddle,
            urlcolor=virdark,
            pdfauthor={Pontus Laurell},
	      pdftitle={Quantifying and controlling entanglement in the quantum magnet Cs2CoCl4}]{hyperref}
\usepackage{verbatim}
\usepackage{ulem}

\begin{document}

%
%

\title{Quantifying and controlling entanglement in the quantum magnet \texorpdfstring{Cs$_2$CoCl$_4$}{Cs2CoCl4}}
\author{Pontus Laurell}
\email{laurell@utexas.edu}
\affiliation{Center for Nanophase Materials Sciences, Oak Ridge National Laboratory, Oak Ridge, Tennessee 37831, USA}
\affiliation{Computational Science and Engineering Division, Oak Ridge National Laboratory, Oak Ridge, Tennessee 37831, USA}
\author{Allen Scheie}
\affiliation{Neutron Scattering Division, Oak Ridge National Laboratory, Oak Ridge, Tennessee 37831, USA}
\author{Chiron J. Mukherjee}
\affiliation{Science Department, Drew School, San Francisco, California 94115, USA}
\affiliation{Clarendon Laboratory, Oxford University, Parks Road, Oxford OX1 3PU, United Kingdom}
\author{Michael M. Koza}
\affiliation{Institut Laue-Langevin, 38042 Grenoble Cedex 9, France}
\author{Mechtild Enderle}
\affiliation{Institut Laue-Langevin, 38042 Grenoble Cedex 9, France}
\author{Zbigniew Tylczynski}
\affiliation{Faculty of Physics, Adam Mickiewicz University, 61-614 Pozna\'n, Poland}
\author{Satoshi Okamoto}
\affiliation{Materials Science and Technology Division, Oak Ridge National Laboratory, Oak Ridge, Tennessee 37831, USA}
\affiliation{Quantum Science Center, Oak Ridge National Laboratory, Tennessee 37831, USA}
\author{Radu Coldea}
\affiliation{Clarendon Laboratory, Oxford University, Parks Road, Oxford OX1 3PU, United Kingdom}
\author{D. Alan Tennant}
\email{tennantda@ornl.gov}
\affiliation{Materials Science and Technology Division, Oak Ridge National Laboratory, Oak Ridge, Tennessee 37831, USA}
\affiliation{Quantum Science Center, Oak Ridge National Laboratory, Tennessee 37831, USA}
\affiliation{Shull-Wollan Center, Oak Ridge National Laboratory, Tennessee 37831, USA}
\author{Gonzalo Alvarez}
\email{alvarezcampg@ornl.gov}
\affiliation{Center for Nanophase Materials Sciences, Oak Ridge National Laboratory, Oak Ridge, Tennessee 37831, USA}
\affiliation{Computational Science and Engineering Division, Oak Ridge National Laboratory, Oak Ridge, Tennessee 37831, USA}
\date{\today}


\begin{abstract}
	The lack of methods to experimentally detect and quantify entanglement in quantum matter impedes our ability to identify materials hosting highly entangled phases, such as quantum spin liquids. We thus investigate the feasibility of using inelastic neutron scattering (INS) to implement a model-independent measurement protocol for entanglement based on three entanglement witnesses: one-tangle, two-tangle, and quantum Fisher information (QFI). We perform high-resolution INS measurements on Cs$_2$CoCl$_4$, a close realization of the $S=1/2$ transverse-field XXZ spin chain, where we can control entanglement using the magnetic field, and compare with density-matrix renormalization group calculations for validation. 
	The three witnesses allow us to infer entanglement properties and make deductions about the quantum state in the material. We find QFI to be a particularly robust experimental probe of entanglement, whereas the one- and two-tangles require more careful analysis. Our results lay the foundation for a general entanglement detection protocol for quantum spin systems.
\end{abstract}
\maketitle

\emph{Introduction.}---%
Quantum entanglement is increasingly considered a vital resource for novel effects and applications \cite{RevModPhys.81.865}. Entanglement is also central to our understanding of many-body systems \cite{RevModPhys.80.517, Vedral2008}, where it forms a deep connection between condensed matter physics and quantum information. Phenomena such as quantum spin liquids \cite{Savary2017},
topological order \cite{RevModPhys.89.041004}, quantum criticality \cite{PhysRevLett.90.227902,Calabrese2004}, and thermalization in quantum systems \cite{RevModPhys.91.021001}, are all inherently related to entanglement properties. It is crucial to develop experimental protocols to detect and quantify entanglement in the solid state, in order to allow unambiguous and rapid identification of quantum materials suitable for new applications, and novel insights into complex quantum phenomena.

Due to the rich structure of many-body states, a number of different entanglement measures have been introduced. The most important example in condensed matter theory is entanglement entropy (EE), used to quantify bipartite entanglement. Yet there is no ``EE observable'' that can be probed directly, which makes experimentally quantifying entanglement in many-body systems challenging \cite{RevModPhys.80.517,Guehne2009}. Although EE has been measured in cold-atom \cite{Islam2015,Kaufman2016} and photonic systems \cite{Pitsios2017}, neither approach is suitable for probing entanglement in macroscopic condensed matter systems.

In special cases, entanglement can be inferred through neutron scattering experiments. For instance, two-spin entanglement within and between dimers \cite{PhysRevA.73.012110,copper_nitrate}, and entanglement between two molecular magnet qubits  \cite{Garlatti2017} have been characterized with neutrons. Also, certain  low-dimensional spin systems can be shown to have entanglement via close comparison with theory \cite{mourigal2013fractional,Piazza2015,Christensen2007}. However, these approaches rely on tractable models, with either small Hilbert spaces or special ground states, which are limited to a handful of systems. For most strongly correlated systems, such methods are not applicable, calling for model-independent approaches.

A promising approach, which we explore in this Letter, is using entanglement witnesses (EWs) \cite{RevModPhys.80.517,Vedral2008,Guehne2009}, i.e. observables that can be used to identify some \emph{subset} of entangled states. We consider (i) one-tangle ($\tau_1$) \cite{PhysRevA.61.052306,PhysRevA.69.022304,PhysRevLett.93.167203}, (ii) concurrence or two-tangle ($\tau_2$) \cite{PhysRevA.61.052306,PhysRevA.69.022304,PhysRevA.73.012110,PhysRevA.74.022322, Garlatti2017}, and (iii) quantum Fisher information (QFI) \cite{PhysRevLett.102.100401,Strobel2014,Hauke2016}. These EWs witness (i) entanglement between a spin and the rest of the system, (ii) pairwise entanglement, and (iii) multipartite entanglement, respectively, and thus provide complementary information. All three EWs are accessible to inelastic neutron scattering (INS) experiments. $\tau_1$ and $\tau_2$ can be obtained from ordered moments and spatial spin-spin correlations \cite{PhysRevLett.93.167203}, while QFI can be expressed as an integral of the dynamical spin structure factor (DSF) \cite{Hauke2016}, $S(k,\hbar\omega)$. This powerful formulation of QFI has been applied to experiments on Heisenberg spin chains \cite{PhysRevResearch.2.043329,Scheie2021}, but remains otherwise largely unexplored.

We contrast EE and the mentioned EWs in the spin-$1/2$ transverse-field XXZ chain. The one-dimensional setting confers an enhanced susceptibility to fluctuations and a higher degree of theoretical tractability, making it an excellent proving ground for our EW protocol. The system hosts two distinct quantum critical points (QCPs) and a classical, minimally entangled point, and thus provides a range of interesting behaviors. We study this model numerically using the density matrix renormalization group (DMRG) \cite{PhysRevLett.69.2863,PhysRevB.48.10345,Alvarez2009}. We also report high-resolution INS data on the chain compound Cs$_2$CoCl$_4$, known to be an excellent realization of the XXZ model \cite{Algra1976,doi:10.1063/1.433962,PhysRevB.28.3904,PhysRevB.65.144432,Mukherjee2004,PhysRevLett.111.187202,PhysRevB.91.024423}. We find that QFI values extracted from experiment and simulation show good agreement, demonstrating it is an experimentally viable probe of entanglement. We also find the experimental one-tangle to deviate from theory in a potentially revealing manner, whereas the two-tangle extraction requires spin-polarization-resolved experiments.

\emph{Transverse-field XXZ chain.}---%
\begin{figure}
	\includegraphics[width=\columnwidth]{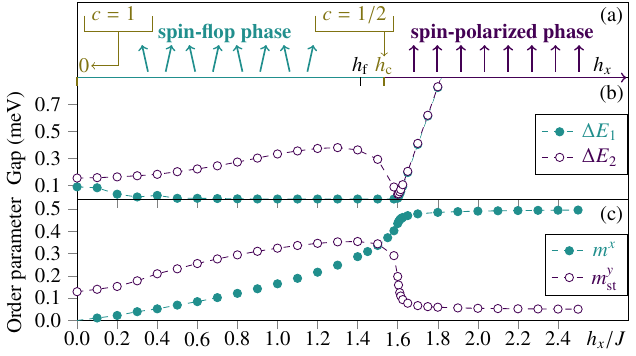}
	\caption{\label{fig:map}
		(a) Schematic phase diagram for $-1 \le \Delta \le 1$. 	There are two quantum critical points at $h_x=0$ and $h_\mathrm{c}$, and a classical point at a factoring field $h_\mathrm{f}$, close to $h_\mathrm{c}$. For $\Delta=0.25$, $h_\mathrm{f}\approx 1.58J$ and $h_\mathrm{c}\approx 1.6J$. The distance $h_\mathrm{c}-h_\mathrm{f}$ is exaggerated for clarity. (b) Energy gap, and (c) magnetization and staggered magnetization \cite{Supplemental} from DMRG for $\Delta = 0.25$, $J=0.23$ meV. Non-vanishing $\Delta E_i$, $m^y_\mathrm{st}$ at $h_x=0$, and $m^y_\mathrm{st} \neq 0$ at $h_x\ge h_\mathrm{c}$, are due to a finite-size effect in the DMRG calculation.}
\end{figure}
A particularly rich yet simple system is found in the XXZ chain,
\begin{equation}
	H	=	\sum_{i=1}^L \left[ J \left( S_i^x S_{i+1}^x + S_i^y S_{i+1}^y + \Delta S_i^z S_{i+1}^z \right) + h_x S_i^x \right],	\label{eq:HXXZ}
\end{equation}
where $S_i^\alpha,\, \alpha\in\{x,y,z\},$ are spin-$1/2$ operators, $\Delta$ represents exchange anisotropy, and $h_x$ is a uniform magnetic field in the transverse ($\hat{x}$) direction. For Cs$_2$CoCl$_4$ we take the parameters $J=0.23$ meV and $\Delta = 0.25$ \cite{PhysRevB.65.144432}, but we note $\Delta \approx 0.12$ has been proposed elsewhere \cite{PhysRevLett.111.187202,PhysRevB.91.024423}.
(We consider other $\Delta$ values in the Supplemental Material (SM) \cite{Supplemental}.) The model is also relevant to quantum simulation using cold atoms in optical lattices \cite{PhysRevX.8.011032}.

For $h_x=0$, the model can be solved exactly using the Bethe ansatz \cite{PhysRev.150.321,PhysRev.150.327,doi:10.1063/1.1705048}. However,  
a finite transverse field breaks integrability, and induces a new source of fluctuations when $\Delta \neq 1$. The model is particularly nontrivial in the spin-flop region, $-1<\Delta < 1$ \cite{Dmitriev2002}, where its phase diagram contains two QCPs, as shown in Fig.~\ref{fig:map}(a). At the first QCP, $h_x=0$, it is equivalent to a gapless Luttinger liquid, which is described by a conformal field theory (CFT) with central charge $c=1$ \cite{PhysRevLett.90.227902}. At $h_x>0$, a gapped, long-range N\'eel order develops, with a staggered magnetization component, $m^y_\mathrm{st}$, along $\hat{y}$, and a magnetization component, $m^x$, along $\hat{x}$ \cite{Supplemental}. This order remains up to a critical field, $h_\mathrm{c}$, where the system is described by a $c=1/2$ CFT. Above $h_\mathrm{c}$ a gapped, nondegenerate spin-polarized (paramagnetic) phase develops and $m^x$ saturates asymptotically. There also exists a ``classical'' or factoring field, $h_\mathrm{f} (\Delta) = J \sqrt{2 (1+\Delta)}$, where the ground state is exactly of the classical spin-flop N\'eel type \cite{Kurmann1982,PhysRevB.32.5845}.  At $h_\mathrm{f}$, quantum fluctuations are precisely balanced by the field, and entanglement estimators indicate an entanglement transition \cite{PhysRevLett.93.167203,Abouie_2010,PhysRevLett.108.240503,PhysRevA.96.052303}.

The model has previously been studied using Jordan-Wigner fermion mean-field theory \cite{Dmitriev2002,PhysRevB.68.134431}, exact diagonalization \cite{PhysRevB.73.054410}, DMRG  \cite{Capraro2002,PhysRevB.68.134431,PhysRevB.94.085136}, and quantum Monte Carlo methods \cite{PhysRevLett.93.167203}. The real-frequency dynamics  were studied in Refs.~\cite{PhysRevB.68.134431,PhysRevB.94.085136}, where the mean-field theory \cite{PhysRevB.68.134431} was found to give qualitatively different spectra to the DMRG calculation \cite{PhysRevB.94.085136} at $h_x\le h_\mathrm{c}$. Here we use a $T=0$ DMRG method described in SM \cite{Supplemental}. Care is taken to relate our finite-size ($L=100$ unless stated otherwise) results in the spin-flop region to the thermodynamic limit \cite{PhysRevB.68.134431,PhysRevB.94.085136,Supplemental}. There is a finite-size gap between a unique ground state and the first excited state, $\Delta E_1=E_1-E_0$, where $E_n$ is the energy of the $n$th state. The physical excitation gap is given instead by $\Delta E_2=E_2-E_0 > \Delta E_1$, as shown in Fig.~\ref{fig:map}(b). Magnetization is plotted in Fig.~\ref{fig:map}(c).

\emph{Experimental method.}---%
INS data on a high-quality $9$~g solution-grown Cs$_2$CoCl$_4$ single crystal were collected using the direct-geometry time-of-flight spectrometer IN6 at Institut Laue-Langevin, with monochromatic incident neutrons of $2.35$~meV. Cooling was provided by a dilution refrigerator, and data was collected at $70$~mK ($\approx 0.026 J \approx 6~\mu$eV). The sample was oriented with crystallographic $b,c$-axes in the horizontal scattering plane. Magnetic fields up to $2.5$~T were applied along the $a$-axis using a vertical field cryomagnet. For more details about the experiments, see Ref.~\cite{chiron_phd}. Raw neutron counts  were normalized by the integrated quasielastic incoherent scattering to account in a first approximation for neutron absorption from the sample. The nonmagnetic background was modeled and subtracted, and resulting counts were divided by the squared spherical magnetic form factor for Co$^{2+}$, so resulting intensities are proportional to the purely magnetic scattering cross section.

The Co$^{2+}$ ions in Cs$_2$CoCl$_4$ form a Kramers doublet, which can be described by an effective spin $S=1/2$. Magnetic interactions between Co$^{2+}$ ions are quasi-1D along the $b$-axis, with exchange interaction much lower than the energy gap to higher crystal field levels, resulting in an effective spin-1/2 Hamiltonian with strong XXZ anisotropy. Finite 3D interchain couplings (estimated to be at least an order of magnitude smaller than $J$ \cite{PhysRevB.28.3904,Chatterjee2003,Dmitriev2004,PhysRevB.91.024423}) stabilize long-range order below $T_N=0.212$ K with ordering wavevector $\bm{q}=(0,1/2,1/2),$ where spins point near the $b$-axis. Transverse magnetic fields applied along the $a$-axis suppress this order at $h_\mathrm{c}^\mathrm{exp}=2.10(4)$ T \cite{PhysRevB.65.144432}. This field direction is at an $\approx 40^{\circ}$ angle to the $xy$ easy plane of the spins. This angle---along with interchain couplings \cite{Dmitriev2004,PhysRevB.70.144414}---is expected to renormalize transition fields compared to the in-plane field case considered in Eq.~\eqref{eq:HXXZ}, but not to change the qualitative content of the phase diagram. To compare experimental and DMRG results, we scale fields such that $h_\mathrm{c}^\mathrm{DMRG}\approx 1.604J=h_\mathrm{c}^\mathrm{exp}$.

\begin{figure}
	\includegraphics[width=\columnwidth]{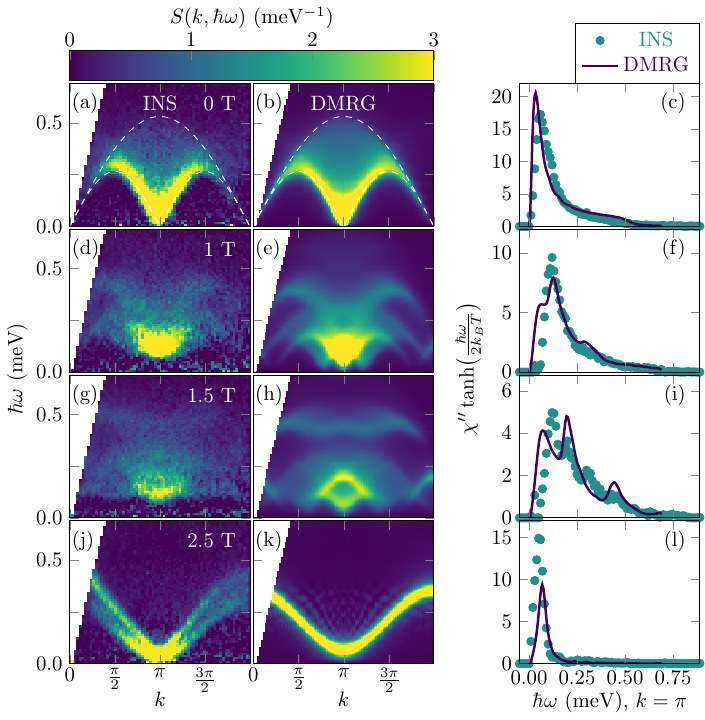}
	\caption{\label{fig:spectrum:XXZ}
		Left column: INS spectra for Cs$_2$CoCl$_4$ at four representative field strengths. Center column: Calculated spectra for the XXZ chain at matching fields, accounting for the experimental polarization factor. Right column: QFI integrand at $k=\pi$. White dashed lines in (a),(b) bound the two-spinon continua. Throughout we designate the wavevector component $k$ along the chain in units of $1/b$.}
\end{figure}
\begin{figure}
	\includegraphics[width=\columnwidth]{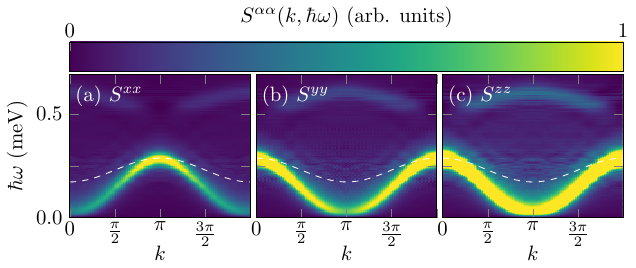}
	\caption{\label{fig:spectrum:XXZ:hf}
		Theoretical DSF for $J=0.23$ meV, $\Delta = 0.25$, and $h_x=1.58114J\approx{}2.07~\text{T}\approx{}h_\mathrm{f}$. Dashed white lines show linear spin wave energy predictions \cite{PhysRevB.32.5845}. Agreement with DMRG is excellent at $k=\pi$ in (a), and $k=0$ in (b),(c). Elsewhere in the Brillouin zone the dispersion is significantly modified by anharmonic terms in the full spin-wave Hamiltonian. Note that the experimental polarization factor has not been applied, see \cite[Eq.~(S5)]{Supplemental}.}
\end{figure}
\emph{Spectra.}---%
Figure~\ref{fig:spectrum:XXZ} compares INS spectra for Cs$_2$CoCl$_4$ with spectra calculated for Eq.~\eqref{eq:HXXZ}. For more field strengths and processing details, see SM \cite{Supplemental}. At low fields the data qualitatively agrees with simulations of the ideal chain model, Eq.~\eqref{eq:HXXZ}. Interchain couplings become qualitatively important near $h_\mathrm{c}$, where the field-dependent gap is of similar strength to the interchain exchange. Interchain couplings also produce a band splitting at high fields, as seen in Fig.~\ref{fig:spectrum:XXZ}(j)--(k), whereas the DMRG spectrum reduces to a single magnon branch. Hence, we conclude that Cs$_2$CoCl$_4$ is 1D-like for weak and intermediate fields. Precise modeling of interchain effects is beyond the scope of this work.

At zero field the main contributions to the DSF come from the two-spinon continuum \cite{Muller1981,PhysRevLett.95.077201}, bounds of which are drawn in Fig.~\ref{fig:spectrum:XXZ}(a)--(b). At finite field the excitation branches begin to split [Fig.~\ref{fig:spectrum:XXZ}(d),(e)], eventually decoupling the upper branch from the low-energy excitations, forming a high-energy feature at $\hbar\omega \geq 0.4$ meV. As Ref.~\cite{PhysRevB.94.085136} noted, this feature is beyond the mean-field prediction \cite{PhysRevB.68.134431}. Here we see it is present in the experimental material [panels (d),(g)] and DMRG [panels (e),(h)]. The intensity of this high-energy feature weakens as $h_c$ is approached from below, and as Fig.~\ref{fig:spectrum:XXZ}(j),(k) show, it disappears above $h_c$. To understand its origin, it is instructive to consider 
the factoring field. While the ground state at $h_\mathrm{f}$ is classical, the dynamics cannot be fully described using linear spin-wave theory (LSWT) \cite{PhysRevB.28.1529,PhysRevB.32.5845}. For Eq.~\eqref{eq:HXXZ} the dynamics is LSWT-like only near $k=\pi$ for $S^{xx} \left( k,\hbar\omega\right)$, and near $k=0$ for $S^{yy/zz} \left( k,\hbar\omega\right)$ \cite{PhysRevB.32.5845}. As Fig.~\ref{fig:spectrum:XXZ:hf} shows, this behavior agrees well with DMRG. The high-energy feature vanishes at $k$ points where LSWT is exact, heavily suggesting its origin is in multi-magnon physics, as proposed in Ref.~\cite{PhysRevB.94.085136}.

\begin{figure}
	\includegraphics[width=\columnwidth]{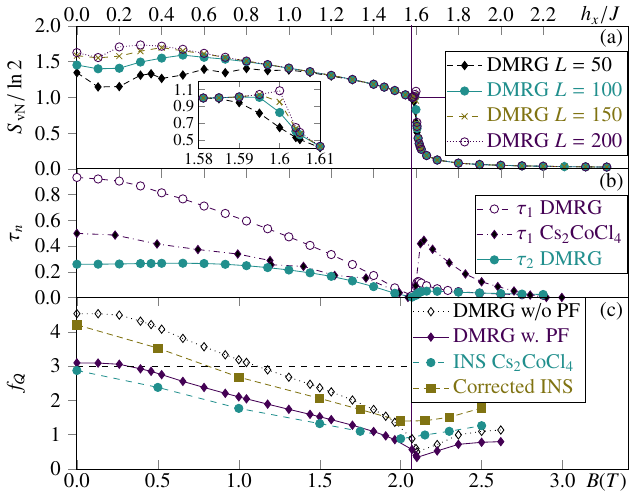}
		\caption{\label{fig:entanglemententropy}
		(a) Entanglement entropy, $S_\mathrm{vN}$, from DMRG as a function of $h_x$. The vertical line indicates the factoring field, where $S_\mathrm{vN}\approx \ln 2$ (horizontal line). For $h_x>h_\mathrm{c}$ there is a steep drop in entropy as the system enters a polarized phase with a non-degenerate ground-state. The inset shows EE near $h_\mathrm{c}$. (b) The approximate one- ($\tau_1$) and two-tangles ($\tau_2$) reach a minimum at $h_\mathrm{f}$. (c) QFI from INS and DMRG $S(k,\hbar\omega)$. Above the horizontal dashed line QFI indicates the presence of \emph{at least} bipartite entanglement. Below it QFI cannot be used to distinguish separable and entangled states. The polarization factor (PF)-corrected INS $f_\mathcal{Q}$ line is obtained by scaling $f_\mathcal{Q}^\mathrm{INS}$ by the ratio between the two DMRG $f_\mathcal{Q}$ values \cite{Supplemental}.
	}
\end{figure}
\emph{Entanglement.}---%
We now investigate the quantum phase transition (QPT) of Eq.~\eqref{eq:HXXZ} and Cs$_2$CoCl$_4$ using entanglement measures. Figure~\ref{fig:entanglemententropy}(a) shows half-chain von Neumann EE. At QCPs, in a system of length $L$ with open boundaries, it is expected to follow the CFT expression  \cite{Calabrese2004}, $S_\mathrm{vN}	=	\frac{c}{6} \ln \left[ \frac{L}{\pi} \right] + C,	\label{eq:entropy:finite}$ where $C$ is a non-universal correction. We observe approximately logarithmic scaling at the QCPs, and saturation of $S_\mathrm{vN}$ for most non-critical fields \cite{PhysRevLett.90.227902}. Notably, we find at $h_\mathrm{f}$ that $S_{vN}=\ln 2$ to good approximation, consistent with a two-fold classical ground-state degeneracy.

Another sharp signature of the classical state has previously been found using entanglement estimators \cite{PhysRevLett.93.167203,PhysRevLett.108.240503,PhysRevA.96.052303}. We consider one-tangle, $\tau_1$, which quantifies entanglement between a single site and the rest of the system, and two-tangle, $\tau_2$, which quantifies the total pairwise entanglement in the system, and satisfies $\tau_2<\tau_1$ \cite{PhysRevA.61.052306,PhysRevLett.96.220503}. For translation-invariant $S=1/2$ systems, $\tau_1$ can be defined in terms of spin expectation values at a given site $j$, $\tau_1	=	1-4\sum_\alpha \left( \langle S_j^\alpha\rangle\right)^2$. It is useful for interpreting experiments, with the caveat that it is only strictly defined at $T=0$. We approximate $\tau_1$ by keeping only ferro- and antiferromagnetic ordered moments \cite{Supplemental}. The theoretical prediction is shown in Fig.~\ref{fig:entanglemententropy}(b), along with experimental results obtained \cite{Supplemental} using $80$ mK ($\approx 0.03J$) data from Ref.~\cite{PhysRevB.65.144432}. At zero field the experimental $\tau_1$ is reduced from the theoretical value due to magnetic ordering at low temperature, but still indicates substantial entanglement. We discuss the high $\tau_1$ at $B>2$~T later. 

Next, two-tangle is defined as $\tau_2=2\sum_{r\neq 0}C_{r}^2,$ where $C_r$ is the concurrence for separation $r$. $C_r$ for the $S=1/2$ XXZ model absent spontaneous symmetry breaking ($m^y_\mathrm{st}=0$) can be defined \cite{PhysRevA.69.022304, PhysRevLett.93.167203,PhysRevA.74.022322,Abouie_2010} $C_{r} =	2\max \left\{ 0, C_{r}', C_{r}'' \right\},$ where
\begin{align}
C_{r}'	&=	\left| \langle S_i^y S_{i+r}^y\rangle + \langle S_i^zS_{i+r}^z\rangle \right| - \sqrt{\left( \frac{1}{4} + \langle S_i^xS_{i+r}^x\rangle \right)^2 - \left( m^x\right)^2},\\
C_{r}''	&=	\langle S_i^xS_{i+r}^x\rangle + \left| \langle S_i^y S_{i+r}^y\rangle - \langle S_i^zS_{i+r}^z\rangle \right| - \frac{1}{4}.
\end{align}
This definition acts as a lower bound for pairwise entanglement in the symmetry-broken state \cite{PhysRevA.68.060301, PhysRevLett.97.257201}. 
While such correlation functions are straightforward to compute theoretically, for anisotropic systems they require spin-polarization-resolved techniques to measure experimentally. Since we have not conducted such experiments, we plot only the theoretical $\tau_2$ in Fig.~\ref{fig:entanglemententropy}(b). (In \cite{Supplemental} we simulate a polarized INS experiment by using DMRG to correct for polarization factors (PFs), and estimate concurrence and $\tau_2$ from unpolarized data. We find rough agreement between experiment and theory at low fields, suggesting $\tau_2$ could be a reliable EW with carefully performed experiments.)

Finally, we come to the quantum Fisher information. The QFI density, $f_\mathcal{Q}$, can be expressed \cite{Hauke2016}
\begin{equation}
f_\mathcal{Q}({k,T}) = \frac{4}{\pi} \int_0^{\infty} \mathrm{d} (\hbar \omega) \tanh \left( \frac{\hbar \omega}{2 k_B T} \right) \chi\prime\prime (k, \hbar \omega, T) \label{eq:QFI},
\end{equation}
where the dynamical susceptibility, $\chi\prime\prime$, is linked to the DSF through the fluctuation-dissipation theorem, $\chi\prime\prime\left(k,\hbar\omega,T\right)=\tanh\left(\sfrac{\hbar\omega}{2k_BT}\right)S\left(k,\hbar\omega\right)$, and $S(k,\hbar\omega)$ is normalized per site (i.e. intensive) according to the the sum rule 
$\sum_{\alpha\in\{x,y,z\}}\int_{-\infty}^{\infty} \mathrm{d}(\hbar\omega) \int_{0}^{2\pi}
 \mathrm{d}k \, S^{\alpha\alpha} \left(k, \hbar\omega\right) =S(S+1).$
We are interested in $f_\mathcal{Q} (k=\pi)$, which witnesses entanglement associated with the AFM ordering \cite{Supplemental}. Importantly, one can derive bounds for $f_\mathcal{Q}$ that can only be met by certain classes of entangled states \cite{PhysRevA.85.022321,PhysRevA.85.022322,Pezze2014}. For unpolarized neutron scattering and $S=1/2$ systems, $f_\mathcal{Q}>3n$, with $n$ a divisor of $L$, indicates the system is \emph{at least} $n+1$-partite entangled \cite{Supplemental}.

Figure ~\ref{fig:entanglemententropy}(c) shows QFI determined from INS data normalized against DMRG \cite{Supplemental}, and from DMRG with and without PF applied. All QFI integrals used $T=70$~mK. In all cases, maximal $f_\mathcal{Q}$ occurs at $h_x=0$. Unlike $\tau_1$, QFI is insensitive to zero-field magnetic order since elastic peaks are suppressed by the $\tanh$ factor in Eq.~\eqref{eq:QFI}. The results indicate the experimental PF suppresses QFI below the lower bound required to observe bipartite entanglement. Using DMRG intensities we can obtain PF-corrected values \cite{Supplemental}, which do witness at least bipartite entanglement at the lowest measured fields. This finding highlights that it is easy to underestimate the underlying QFI due to resolution effects, and calls for higher resolution in future experiments. Additionally, it would be valuable to derive tighter bounds on $f_\mathcal{Q}$, even if they do not apply in general \cite{PhysRevA.95.032330,Almeida2020}.

There is qualitative and reasonably quantitative agreement between DMRG and INS QFI at intermediate fields ($\lessapprox 1.75$ T), but not at high fields, where interchain coupling causes deviations from ideal 1D behavior. In particular, interchain coupling raises the field required for full polarization, which may explain the observed increase in $f_\mathcal{Q}$ above $h_\mathrm{c}$. As $h_x\rightarrow \infty$ we expect $f_\mathcal{Q}$ to vanish. In addition, the $f_\mathcal{Q}^\mathrm{DMRG}$ minimum occurs at $h_\mathrm{c}$, while the $f_\mathcal{Q}^\mathrm{INS}$ minimum appears at a lower field, likely due to deviations from ideal 1D behavior.

Another deviation from 1D behavior is seen in the large $\tau_1$ at $B>2$ T [Fig.~\ref{fig:entanglemententropy}(b)], which could naively be interpreted as a sign that the system has entered a highly entangled state. However, this scenario seems unlikely given the observed $f_\mathcal{Q}$ behavior, and suppression of magnetic fluctuations at high field. Instead, $\tau_1$ is likely overestimated in this region due to spin correlations not captured by the ferro- and antiferromagnetic ordered moments used to evaluate $\tau_1$, induced by the small, but finite interchain couplings, which become relevant for fields near $h_\mathrm{c}$ where the spin gap is small and the system is near-critical. Capturing such 3D effects is beyond the scope of the current paper.

On its face, the low $f_\mathcal{Q}$ values at $h_\mathrm{c}$ may seem incompatible with the prediction that CFTs have both large bipartite and multipartite entanglement \cite{PhysRevD.96.126007}. QFI near $h_\mathrm{c}$ is low because the N\'eel order parameter becomes vanishingly small near the transition to the polarized state, such that there is little spectral weight available for entanglement \cite{Blanc2018}, and so $f_\mathcal{Q}(k=\pi)$ is not an effective witness at this QCP. We generically expect QFI associated with antiferromagnetic ordering vectors to demonstrate significant entanglement only away from paramagnetic transitions. This illustrates a general limitation of EWs: they are not universal \cite{Guehne2009}. Thus additional EWs would be required to experimentally characterize the entanglement properties of the transverse-field XXZ chain in the entire field range.

\emph{Conclusion.}---%
We have contrasted several entanglement measures by applying them to Cs$_2$CoCl$_4$ and the transverse-field XXZ chain. Although we are unable to directly witness genuine multipartite entanglement in Cs$_2$CoCl$_4$, the strong agreement between DMRG and INS QFI suggests QFI is already a useful tool for qualitative investigations of entanglement properties, potentially even for topological phases \cite{PhysRevLett.119.250401,Gabbrielli2018,PhysRevB.102.224401}. With improved resolution and bounds it can also prove valuable for directly quantifying \emph{local} entanglement in materials. QFI can be used in combination with other EWs, to infer entanglement properties as a control parameter is tuned. Such combinations may be required since paramagnetic QPTs remain inaccessible to $f_\mathcal{Q}(k=\pi)$.
We find both one-tangle and QFI to be useful measures for inelastic experiments in general, while two-tangle requires polarization analysis. 
We expect the model-independent approach we outline here, which applies to many spectroscopic techniques and also to higher-dimensional systems, will prove useful in identifying materials with entangled states and highly quantum properties. 
As the search for materials realizing exotic quantum states continues, EWs can allow clear discrimination between truly entangled, and disordered non-entangled states.

\section*{Acknowledgements} %
\begin{acknowledgments}
	The research by PL, SO, and GA was supported by the scientific Discovery through Advanced Computing (SciDAC) program funded by U.S. Department of Energy, Office of Science, Advanced Scientific Computing Research and Basic Energy Sciences, Division of Materials Sciences and Engineering. GA was in part supported by the ExaTN ORNL LDRD. The work by DAT is supported by the Quantum Science Center (QSC), a National Quantum Information Science Research Center of the U.S. Department of Energy (DOE). AS was supported by the DOE Office of Science, Basic Energy Sciences, Scientific User Facilities Division. Software development has been partially supported by the Center for Nanophase Materials Sciences, which is a DOE Office of Science User Facility. RC acknowledges support from the European Research Council under the European Union Horizon 2020 Research and Innovation Programme via Grant Agreement 788814-EQFT. Access to the data reported in this paper will be made available from Ref.~\cite{Data}.
\end{acknowledgments}


%

\end{document}


\title{Supplemental Material for\texorpdfstring{\\}{}``Quantifying and controlling entanglement in the quantum magnet \texorpdfstring{Cs$_2$CoCl$_4$}{Cs2CoCl4}''}

\author{Pontus Laurell}
\email{laurellp@utexas.edu}
\affiliation{Center for Nanophase Materials Sciences, Oak Ridge National Laboratory, Oak Ridge, Tennessee 37831, USA}
\affiliation{Computational Science and Engineering Division, Oak Ridge National Laboratory, Oak Ridge, Tennessee 37831, USA}
\author{Allen Scheie}
\affiliation{Neutron Scattering Division, Oak Ridge National Laboratory, Oak Ridge, Tennessee 37831, USA}
\author{Chiron J. Mukherjee}
\affiliation{Science Department, Drew School, San Francisco, California 94115, USA}
\affiliation{Clarendon Laboratory, Oxford University, Parks Road, Oxford OX1 3PU, United Kingdom}
\author{Michael M. Koza}
\affiliation{Institut Laue-Langevin, 38042 Grenoble Cedex 9, France}
\author{Mechtild Enderle}
\affiliation{Institut Laue-Langevin, 38042 Grenoble Cedex 9, France}
\author{Zbigniew Tylczynski}
\affiliation{Faculty of Physics, Adam Mickiewicz University, 61-614 Pozna\'n, Poland}
\author{Satoshi Okamoto}
\affiliation{Materials Science and Technology Division, Oak Ridge National Laboratory, Oak Ridge, Tennessee
\affiliation{Quantum Science Center, Oak Ridge National Laboratory, Tennessee 37831, USA}37831, USA}
\author{Radu Coldea}
\affiliation{Clarendon Laboratory, Oxford University, Parks Road, Oxford OX1 3PU, United Kingdom}
\author{D. Alan Tennant}
\email{tennantda@ornl.gov}
\affiliation{Materials Science and Technology Division, Oak Ridge National Laboratory, Oak Ridge, Tennessee 37831, USA}
\affiliation{Quantum Science Center, Oak Ridge National Laboratory, Tennessee 37831, USA}
\affiliation{Shull-Wollan Center, Oak Ridge National Laboratory, Tennessee 37831, USA}
\author{Gonzalo Alvarez}
\email{alvarezcampg@ornl.gov}
\affiliation{Center for Nanophase Materials Sciences, Oak Ridge National Laboratory, Oak Ridge, Tennessee 37831, USA}
\affiliation{Computational Science and Engineering Division, Oak Ridge National Laboratory, Oak Ridge, Tennessee 37831, USA}
\date{\today}

\def\theequation{S\arabic{equation}}
\def\thefigure{S\arabic{figure}}

\maketitle

In this supplement we (i) describe the numerical method and data analysis techniques in more detail, (ii) provide additional experimental and numerical results, (iii) explain how to reproduce the numerical results, and (iv) collect useful analytical results about the 1D spin model.

\section{Methods}

\subsection{Numerical method}

\begin{figure}
	\includegraphics[width=\columnwidth]{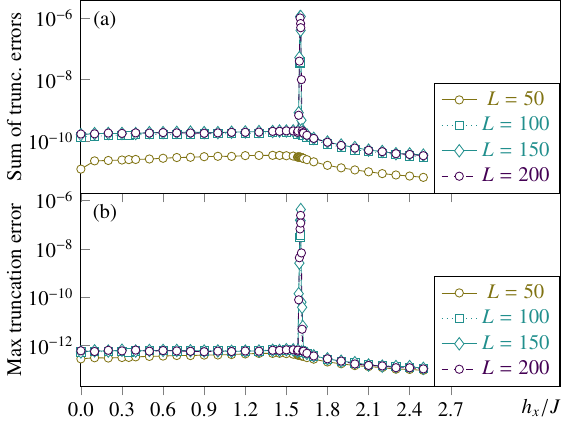}
	\caption{\label{fig:supp:truncation}Truncation errors. (a) Sum of truncation errors and (b) maximal single truncation error in DMRG ground state runs. The peak occurs near the critical field $h_\mathrm{c}$. Excellent numerical accuracy is achieved at other fields.}
\end{figure}

The numerical calculations used the DMRG++ program \cite{Alvarez2009}. Unlike Ref.~\cite{PhysRevB.94.085136}, which applied time-dependent DMRG to the transverse-field XXZ chain, we work directly in frequency using the Krylov correction vector algorithm \cite{PhysRevB.60.335, PhysRevE.94.053308} to calculate $T=0$ dynamical spin structure factors (DSFs). 
Chains with open boundary conditions (OBC) were used throughout the calculations. For the ground state runs used (i) to calculate entanglement entropy and (ii) as the starting point for dynamics runs, we targeted individual truncation errors of $10^{-10}$ or smaller, while keeping up to $m=1000$ states in the infinite and finite loops for $L=50$ and $L=100$. For $L=150$ and $L=200$, $m=1500$ and $m=2000$ were used, respectively. The resulting truncation errors are plotted in  Fig.~\ref{fig:supp:truncation}.
In contrast, in order to accurately calculate energy gaps and the staggered magnetization, separate calculations were carried out keeping the three states with lowest energy and using $m=1300$ for $h_x\le 1.604J$, except for $h_x=h_\mathrm{f}$. For $h_x=h_\mathrm{f}$ and $h_x> 1.604J$ we instead used $m=1000$ and explicit reorthogonalization at each step, to avoid Lanczos ghost states. This kept the sum of truncation errors below $10^{-6}$ for all fields.

Special care needs to be taken to relate the finite-size DMRG results to the thermodynamic limit \cite{PhysRevB.68.134431,PhysRevB.94.085136}. For $0 < h_x < h_\mathrm{c}$ and $L\rightarrow \infty$, the system has two degenerate N\'eel states. 
For $L$ finite, there is a finite-size gap between a unique ground state and the first excited state, $\Delta E_1=E_1-E_0$, where $E_n$ is the energy of the $n$th state. The physical excitation gap is given instead by $\Delta E_2=E_2-E_0 > \Delta E_1$. The magnetization $m_x$ is defined
\begin{align}
m^x	&=	\langle \psi_0 | S^x_i | \psi_0\rangle,     \label{eq:supp:mx}
\end{align}
where $|\psi_0\rangle$ is the ground state. 
For $L$ finite $\langle\psi_0|S_i^y|\psi_0\rangle=0$ for all sites $i$, and the staggered magnetization is instead estimated using the overlap \cite{PhysRevB.68.134431,PhysRevB.94.085136}
\begin{align}
m_\mathrm{st}^y	&=	\langle \psi_0 | \left(-1\right)^i S_i^y | \psi_1 \rangle,  \label{eq:supp:myst}
\end{align}
where $|\psi_1\rangle$ is the first excited state. 
The reason for this choice is that the more conventional expectation value $\langle \psi_0 | S_i^y | \psi_0\rangle = 0$ for all sites $i$ in a translation-symmetric system of finite volume and with finite-size energy gap $\Delta E_1$. A proof of this statement is given in Mathematical Note IV below. As the thermodynamic limit is approached for $0\le h_x \le h_\mathrm{c}$, the gap $\Delta E_1\rightarrow 0$ resulting in a two-fold ground state degeneracy. In this limit spontaneous symmetry breaking can occur, allowing for finite staggered magnetization in the ground state $|\psi_0(L\rightarrow \infty)\rangle$, which is a linear combination of $|\psi_0\rangle$ and $|\psi_1\rangle$.

The dynamical spin structure factors were calculated using the Krylov correction vector algorithm \cite{PhysRevB.60.335, PhysRevE.94.053308}. The correction vector (CV) approach is computationally expensive compared to methods based on time-dependent DMRG and MPS-based Chebyshev expansion techniques \cite{PhysRevB.94.085136}, as it requires individual calculations at each frequency $\omega$. The advantage of the CV approach is that it allows for constant resolution in frequency space, limited only by the finite Lorentzian broadening $\eta\sim 1/L$. We have used $\eta = 0.1 J$. In particular, the CV approach can provide high frequency resolution near $\omega = 0$ --- here we use $\Delta \omega=0.025J$. The dynamical correlations are calculated in real space, and then Fourier transformed to momentum space using the center-site approximation. The approximation is exact in the thermodynamic limit, but introduces some ``ringing'' in finite systems when there is significant weight at low frequencies. To account for this fact, and for the presence of significant elastic peaks in $S^{xx} \left( k, \omega\right)$ and $S^{yy} \left( k, \omega\right)$ (see Mathematical Notes VI and VII below) we subtract a Lorentzian from $S^{\alpha\alpha}(k,\omega)$ at each momentum $q$ with the same width $\eta$, and with height equal to $S^{\alpha\alpha}(k,\omega=0)$, where $\alpha\in \{x,y,z\}$, before plotting DMRG spectra. (Repeated indices do not imply summation. See also next section for our procedure to normalize the intensity before calculating QFI.) This removes the elastic peaks and (most of) the ringing. Note that, in a numerical calculation of $S(k,\omega)$ with energy broadening and finite frequency resolution, removing the $\omega=0$ Lorentzian can potentially cut off some inelastic features at small but finite $\omega$. In particular, we are unable to identify small gaps in the dynamical correlation functions. However, the QFI integral is itself insensitive to the classical contribution at $\omega=0$ \cite{Hauke2016}.

To minimize the impact of boundary effects, the concurrence and two-tangle calculations excluded the first $5$ and last $5$ sites in the chain in the averaging of spin-spin correlators and on-site magnetization. In contrast, the magnetization and one-tangle data plotted in the main text was averaged over all sites. The DMRG one-tangle was approximated by
\begin{align}
	\tau_1	&=	1-4\sum_\alpha \left( \langle S_j^\alpha\rangle\right)^2	\approx 1-4\left[ 
	\left( m^x \right)^2 + \left( m^y_\mathrm{st}\right)^2 \right],	\label{eq:tau1}
\end{align}
with $m^x$ and $m^y_\mathrm{st}$ given by Eqs.~\eqref{eq:supp:mx}, \eqref{eq:supp:myst} above. This matches the theoretical behavior of the transverse-field XXZ chain in the thermodynamic limit, where e.g. $\langle S_j^z\rangle$ vanishes. It also gives us a more direct comparison with the experimental one-tangle described below. We note that Fig.~\ref{fig:magnetizationComponents} shows $m^x_\mathrm{st}<0.06<m^y_\mathrm{st}$ at zero field and unphysical $0<m^z_\mathrm{st}<0.04 \ll m^x$ at high fields. Thus $m^{x}_\mathrm{st}$ and $m^z_\mathrm{st}$ would only produce negligible contributions to $\tau_1$, making this a good approximation.

\subsection{Intensity normalization procedure}
\begin{figure}
	\includegraphics[width=\columnwidth]{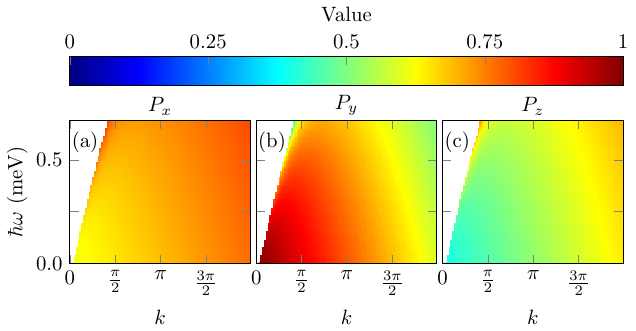}
	\caption{\label{fig:polarization}Experimental polarization factor. At $k=\pi$ the factor preferentially picks out the $S^{xx}$ and $S^{yy}$ components of the DSF over the $S^{zz}$ component.}
\end{figure}

To calculate QFI we need to know the absolute intensity of $S(k,\hbar\omega)$. This was not directly determined in the Cs$_2$CoCl$_4$ experiment, so we have to normalize against the DMRG data. This is complicated by the elastic features mentioned in the previous section. At finite field, they represent a combination of Bragg peaks and an unphysical finite-size artifact. However, at zero field they are entirely due to the finite-size artifact since there is no static long range order in the 1D model. Thus we adopted the following procedure:

\begin{enumerate}
	\item Remove unphysical artifact from zero-field DMRG scattering.
	\item Normalize remaining intensity in components $S^{xx}$, $S^{yy}$, $S^{zz}$, such that the sum satisfies the total moment sum rule, \begin{align} \sum_{\alpha\in\{x,y,z\}}\int_{-\infty}^{\infty} \mathrm{d}(\hbar\omega) \int_{0}^{2\pi}
	\mathrm{d}k \, S^{\alpha\alpha} \left(k, \hbar\omega\right) &=S(S+1).  \label{eq:sumrule}
	\end{align}
	\item Apply experimental polarization factor $P_\alpha(k,\hbar\omega)$ (see Fig.~\ref{fig:polarization}) and energy broadening, to obtain fully processed zero-field DMRG spectra.
	\item Take experimental zero-field spectrum, and normalize inelastic scattering against processed DMRG intensity, accounting for experimental ordered moment (which is in the $xy$-easy plane and has magnitude $1.7\>\mu_\mathrm{B}$ \cite{PhysRevB.65.144432}. Using $g_{xy}=4.8$ \cite{PhysRevB.65.144432}, this means $(\mu/g_{xy})^2/(S(S+1)) = 16.7 \%$ of scattering is elastic).
	\item Use this absolute intensity scale also at finite fields, and to normalize finite-field DMRG spectra.
\end{enumerate}

\begin{figure*}
	\includegraphics[height=\textheight]{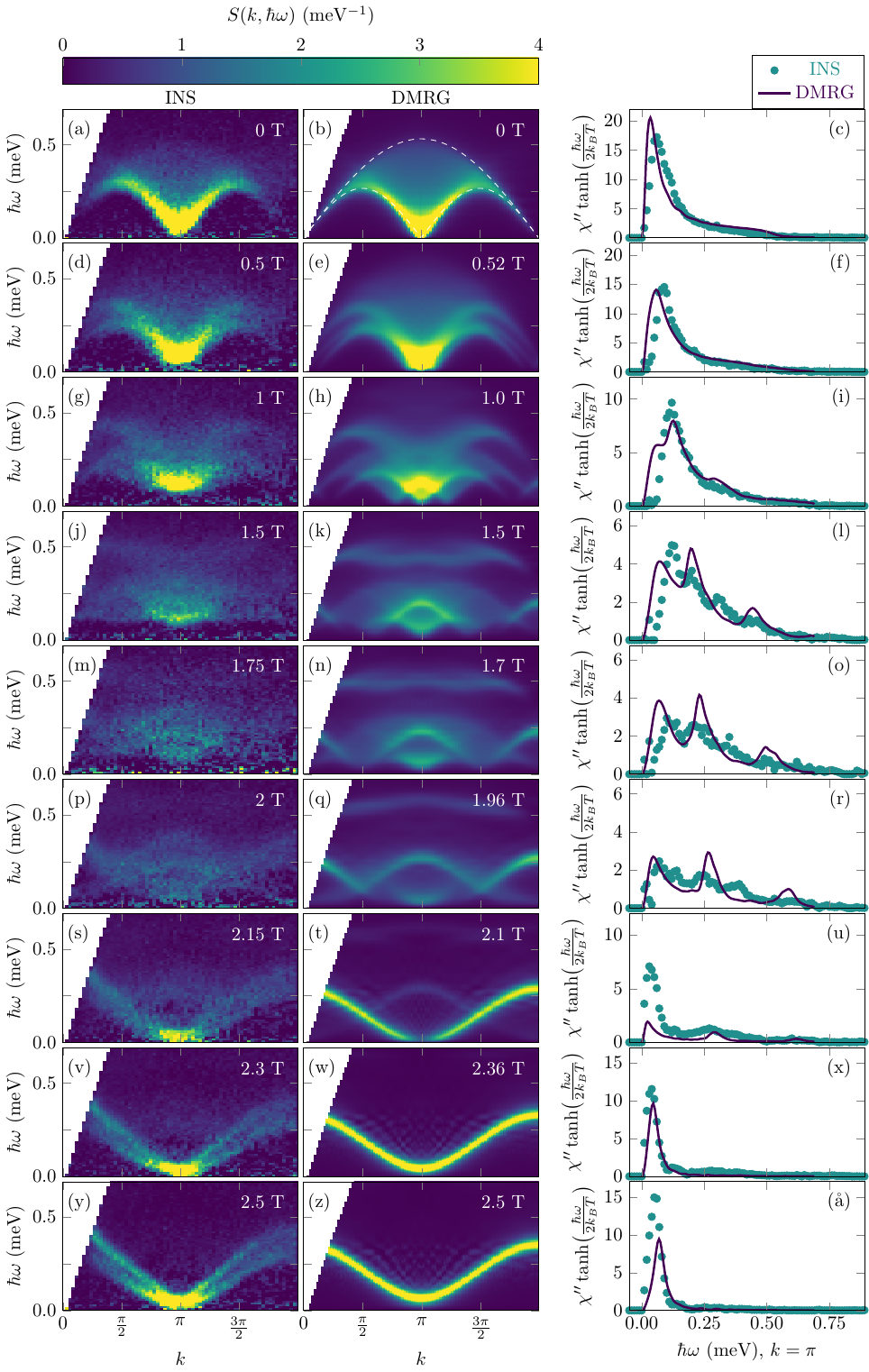}
	\caption{\label{fig:exptvsdmrg}Field evolution of neutron scattering for Cs$_2$CoCl$_4$. Left column: Form factor-corrected experimental spectra. Middle column: Spectra from DMRG, with experimental polarization factor and broadening applied. Right column: QFI integrand at $k=\pi$.}
\end{figure*}

Fig.~\ref{fig:exptvsdmrg} shows the full field evolution of the INS data, and DMRG results at matching fields, with the above normalization procedure applied.

\subsection{Quantum Fisher information notes}
\begin{figure*}
	\includegraphics[width=\textwidth]{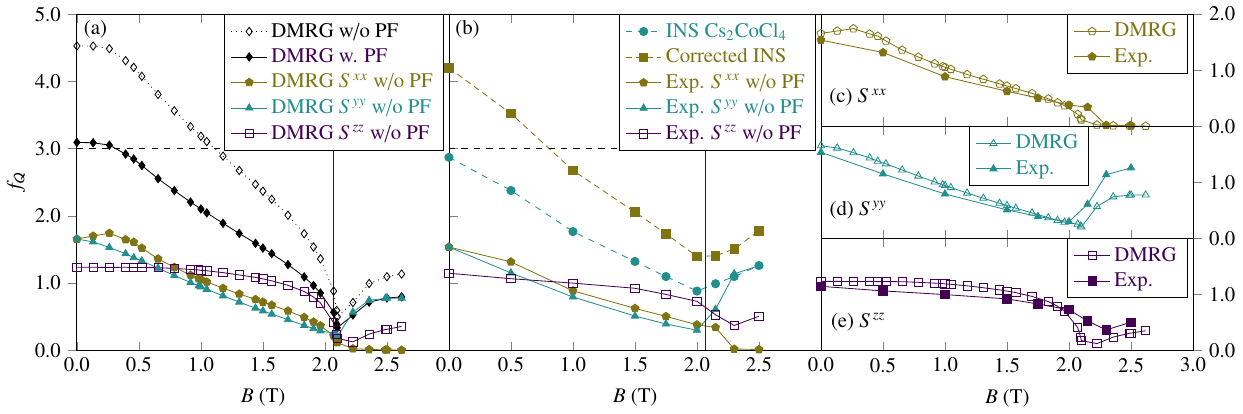}
	\caption{\label{fig:QFIcomponents}Contributions to the quantum Fisher information from individual components of the DSF. (a) shows the DMRG results, and (b) shows the estimated experimental components. Panels (c)--(e) contrast the calculated and estimated experimental components. Above the dashed horizontal line in (a),(b) QFI indicates the presence of at least bipartite entanglement.
	}
\end{figure*}
As mentioned in Ref.~\cite{Hauke2016}, the QFI can be defined for any Hermitian operator $\mathcal{O}$, and expressed as an integral over the dynamical susceptibility associated with $\mathcal{O}$,
\begin{align}
f_\mathcal{Q}({k,T}) &= \frac{4}{\pi} \int_0^{\infty} \mathrm{d} (\hbar \omega) \tanh \left( \frac{\hbar \omega}{2 k_B T} \right) \chi\prime\prime (k, \hbar \omega, T) \label{eq:QFI}.
\end{align}
The interpretation of QFI as a bound for multipartite entanglement is known to apply if $\mathcal{O}$ is a sum of local operators, $\mathcal{O}=\sum_j \mathcal{O}_j,$ where $\mathcal{O}_j$ has a bounded spectrum \cite{Pezze2014}. This is the case for the unpolarized INS experiments described in this work, where on each lattice site $j$ we have $\mathcal{O}_j = c_x S_j^x + c_yS_j^y + c_z S_j^z$ and $S_j^\alpha$. 
The entanglement information is then related to dynamical correlation functions $\langle \mathcal{O}_i(t) \mathcal{O}_j(0)\rangle$, which reflects the fact that entanglement is a form of quantum correlation. 

Due to the symmetry of the Hamiltonian in Eq.~(1) of the main text, only diagonal correlations are important. Hence, it is enough to consider the $S^{xx}(k,\hbar\omega)$, $S^{yy}(k,\hbar\omega)$, $S^{zz}(k,\hbar\omega)$ components in the analysis and DMRG calculations. Thus we may write
\begin{align}
S(k,\hbar\omega)	&=	c_x' S^{xx}(k,\hbar\omega) + c_y'S^{yy}(k,\hbar\omega) + c_z' S^{zz}(k,\hbar\omega),   \label{eq:Skw}
\end{align}
where the $c_\alpha'$ are prefactors, and the $S^{\alpha\alpha}(k,\hbar\omega)$ components are normalized according to the sum rule, equation \eqref{eq:sumrule}. If the experimental polarization factor $P$ (see Fig.~\ref{fig:polarization}) is applied, $c_\alpha'=P_\alpha(k,\hbar\omega)$. When the polarization factor (PF) is \emph{not} applied, we use $c_\alpha'=1\, \forall \alpha$ and obtain the theoretical DSF. Note also that the form of Eqs. \eqref{eq:QFI} and \eqref{eq:Skw} implies that the contributions to the QFI from individual $S^{\alpha\alpha}(k,\hbar\omega)$ components are additive.

In general, the bound required to observe $(n+1)$-partite entanglement is \cite{Hauke2016,Pezze2014}
\begin{align}
f_\mathcal{Q}   &> n \left( h_\mathrm{max} - h_\mathrm{min} \right)^2 = n 4 S^2 \left( c_x^2+c_y^2+c_z^2\right),  \label{eq:bound}
\end{align}
where $h_\mathrm{max}$ and $h_\mathrm{min}$ are the largest and smallest eigenvalues of $\mathcal{O}_j$, respectively. The total moment sum rule implies $c_\alpha=1 \, \forall \alpha$ and for a $S=1/2$ systems we then obtain $h_\mathrm{max}=-h_\mathrm{min}=\sqrt{3}/2$ with $f_\mathcal{Q}>3n$, while for {\it e.g.} $\mathcal{O}_j=S^\alpha$ and $S^{\alpha\alpha}\left( k, \hbar \omega \right)$ $f_\mathcal{Q}>n$.

The contributions to the QFI from individual $S^{\alpha\alpha}(k,\hbar\omega)$ components in the DMRG calculations are shown in Fig.~\ref{fig:QFIcomponents}(a). 
Given the close agreement between DMRG and experiment (at least at low and intermediate fields), we have also estimated the PF-corrected INS $f_\mathcal{Q}$ (plotted in Fig.~4 of the main text) using the DMRG values. The PF-corrected values were obtained as 
\begin{align}
    f_\mathcal{Q}^\text{INS, corrected} &= f_\mathcal{Q}^\text{INS} \frac{f_\mathcal{Q}^\text{DMRG without PF}}{f_\mathcal{Q}^\text{DMRG with PF}}.
\end{align}
Interpolation was used to obtain the DMRG $f_\mathcal{Q}$ values at the experimental fields. Similarly, we estimated the contributions from experimental $S^{\alpha\alpha}(k,\hbar\omega)$ components as
\begin{align}
    f_\mathcal{Q}^{\text{INS }S^{\alpha\alpha}}  &= f_\mathcal{Q}^\text{INS, corrected} \frac{f_\mathcal{Q}^{\text{DMRG }S^{\alpha\alpha}\text{ without PF}}}{f_\mathcal{Q}^\text{DMRG without PF}}.
\end{align}
The resulting estimated QFI contributions are shown in Fig.~\ref{fig:QFIcomponents}(b), and compared with their DMRG counterparts in Figs.~\ref{fig:QFIcomponents} \textbf{c--e}.

At zero field, $S^{xx}(k,\hbar\omega)=S^{yy}(k,\hbar\omega)$ by symmetry, resulting in equal contributions to $f_\mathcal{Q}$. We find two crossovers at finite field. First, at approximately $0.7$ T, $S^{zz}$ becomes the largest contribution to $f_\mathcal{Q}$. The implications on entanglement properties of this crossover are unclear. Second, at approximately $h_\mathrm{c}$, $S^{yy}$ becomes the largest contribution, while $S^{xx}$ approaches zero. The latter effect occurs in the DMRG data because of spin polarization, which develops at lower field than in the experiment. We thus stress that the estimated experimental components are not reliable above $h_\mathrm{c}$, and may be qualitatively wrong. Accurate values could be obtained using polarized neutron scattering experiments. Finally, it is worth noting that, while Fig.~\ref{fig:QFIcomponents} implies that no individual component of the DSF witnesses non-trivial entanglement, this need not be the case for other systems.

In this study we have focused on $k=\pi$ as it corresponds to the antiferromagnetic ordering vector. Other momenta may be of interest in other systems. In general, near a quantum phase transition it is practical to consider $\chi\prime\prime$ associated with an operator $\mathcal{O}_j$ that is relevant in the renormalization group sense, often making the order parameter a good and natural choice \cite{Hauke2016}. For the 1D transverse-XXZ chain, staggered magnetization is a relevant order parameter, suggesting $k=\pi$ is a good choice. In the absence of identified order parameters, one may take the heuristic approach of focusing on the momenta with strongest \emph{inelastic} intensity. These will tend to produce the largest QFI values, and thus are most likely to satisfy the bound necessary to witness entanglement.

\subsection{Extraction of experimental one-tangle}
We use data for the antiferromagnetic (AFM) and ferromagnetic (FM) ordered moments from Ref.~\cite{PhysRevB.65.144432}. Specifically, we take AFM moments from Fig.~12, and FM moments from Fig.~14. The AFM moments are normalized to the absolute moment of the ordered moment at zero field, $m_0=1.7(4)\, \mu_B$ using the easy-plane $g$-factor $g_{xy}=4.8$ \cite{PhysRevB.65.144432}. We obtain $\langle S_0\rangle=m_0/g_{xy}\approx 0.354$ as an overall scale for the AFM moments. The FM moments are normalized using the fact that they are polarized at large fields, such that $\langle S\rangle=0.5$ on the fitted line in Fig.~14 above $\approx 3$ T.

We calculate $\tau_1$ as defined in the text and \eqref{eq:tau1}, equating the FM moment with $m^x$ and the AFM moment with $m^y_\mathrm{st}$, 
interpolating between data points as needed. The data points shown in Fig.~4 of the main text are, for $B\le 2.2$ T and $B>2.9$ T, for field strengths where the AFM moment was measured in Ref.~\cite{PhysRevB.65.144432}. For $2.2 < B < 2.9$ T we have inserted regularly spaced data points assuming a zero AFM ordered moment. For a few data points we find $\tau_1<0$ but $|\tau_1|\approx 0$, whether due to measurement uncertainty or uncertainty in digitizing the plots. In those cases we have set the $\tau_1$ value to zero. We note that this approach accounts only for the FM and AFM ordered moments characterized in Ref.~\cite{PhysRevB.65.144432}, but would not account for other finite static moments. Hence $\tau_1$ may overestimate the entanglement present.

\subsection{Extraction of experimental two-tangle}
\begin{figure*}
	\includegraphics[width=\textwidth]{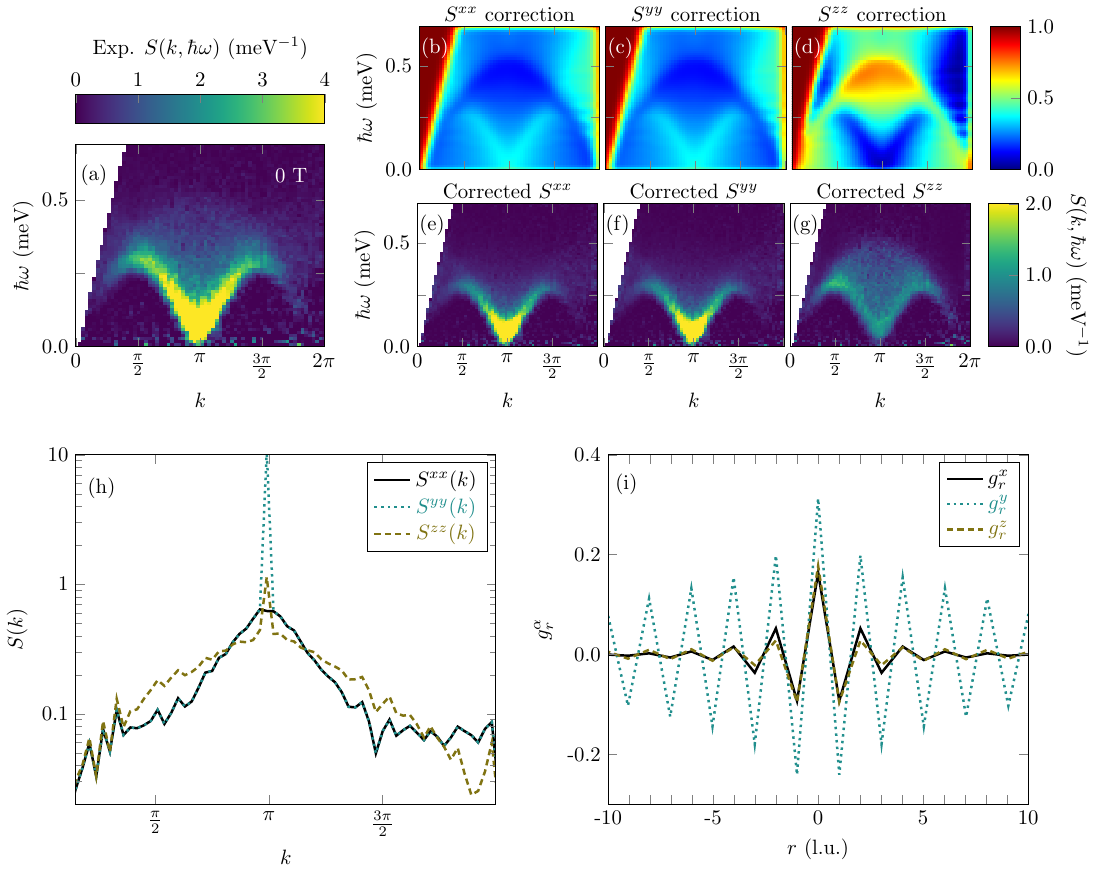}
	\caption{\label{fig:realspace}Extraction of experimental real-space correlation functions. (a) Non-polarized INS spectrum used as starting point. (b)--(d) Correction factors estimated using DMRG intensities. (e)--(g) Approximate PF-corrected polarized scattering components. (h) PF-corrected static spin structure factor. (i) Real-space correlation functions.}
\end{figure*}
\begin{figure}
	\includegraphics[width=\columnwidth]{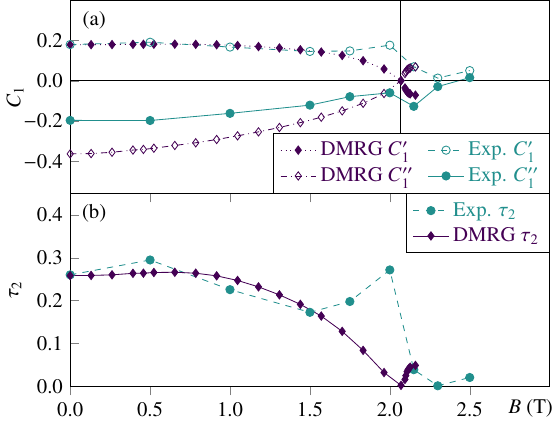}
	\caption{\label{fig:twotangle}Estimated experimental two-tangle and concurrence. (a) Nearest-neighbor concurrence calculated from DMRG and the non-polarized INS experiment. (b) Two-tangle calculated from DMRG and estimated from experiment.}
\end{figure}
We will now describe our approach to obtain the concurrence and $\tau_2$ from the Cs$_2$CoCl$_4$ unpolarized INS data, and its limitations.
To calculate the two-tangle, we need to obtain the real-space equal-time spin-spin correlation functions $g_r^\alpha=\langle S_i^\alpha S_{i+r}^\alpha\rangle$. These are straight-forward to obtain in numerical calculations through ground state expectation values. In contrast, neutron scattering data is generally obtained in reciprocal and frequency space, so a more involved data analysis is required. What is more, obtaining real-space correlation in an anisotropic system requires a polarized inelastic scattering experiment, which was not performed here. 

However, similar to the estimated PF-corrected QFI discussed above, we can obtain approximate PF-corrected polarized components of the inelastic ($\hbar\omega>0$) scattering data by using the DMRG simulated intensity as a correction factor:
\begin{align}
    S^{\alpha\alpha}_\text{INS}(k,\hbar\omega)  &= S_\text{INS}(k,\hbar\omega) \frac{S^{\alpha\alpha}_\text{DMRG without PF}(k,\hbar\omega) }{S^{\alpha\alpha}_\text{DMRG with PF}(k,\hbar\omega)}.
\end{align}
where the correction ratio is calculated pixel-by-pixel. (This expression is correct if DMRG intensity matches the experimental intensity. If discrepancies exist between theory and experiment, this approach is less reliable.)
Next, the elastic scattering is accounted for by adding a Dirac delta function at $k=\pi$ or $k=2\pi$ with height equal to the AFM or FM Bragg intensity from Ref.~\cite{PhysRevB.65.144432}. The components of the PF-corrected static spin structure factor are obtained as
\begin{align}
    S^{\alpha\alpha}(k) &=  \int d(\hbar\omega) \> S^{\alpha\alpha}_\text{INS} (k,\hbar\omega),
\end{align}
and then Fourier transformed to real space. We show an example of this procedure in Fig.~\ref{fig:realspace} using the zero field scattering. At this point, $C_r'$ and $C_r''$ are extracted and $\tau_2$ is calculated using Eqs.~(6--8) in the main text. The results are shown in Fig.~\ref{fig:twotangle}.

The nearest-neighbor concurrence in Fig.~\ref{fig:twotangle}(a) shows good agreement between DMRG and experiment at low and intermediate fields, particularly for $C_1'$. At field strengths close to $h_\mathrm{c}$, the experimental values begin to qualitatively disagree with the DMRG results. We note that this disagreement is consistent with deviations from the 1D model, and the fact that the PF-correction approximations are unreliable at higher fields. However, this implies in particular that we are not able to experimentally observe the entanglement transition \cite{PhysRevA.74.022322} that is evident in the DMRG result where $C'$ and $C''$ both change sign at $h_\mathrm{f}$. At this transition $C'$ and $C''$ also becomes long-ranged.

The limitation from the PF-correction could be overcome in polarized neutron scattering experiments, allowing, in principle, for reliable extraction of real-space correlation functions and concurrences. However, the spatial spin-spin correlations are very sensitive to the low-energy scattering, and assumptions made about it in the data analysis do influence the results. This fact makes the two-tangle a very fragile quantity that is difficult to extract accurately. 

As shown in Fig.~\ref{fig:twotangle}(b) we find that---for low fields, where the polarization factor correction is valid---the experimental $\tau_2$ matches the theoretical $\tau_2$ reasonably well. 
Although these experimental values used a simulated DMRG data set to extract polarization components, they demonstrate that $\tau_2$ can in principle be extracted from polarized inelastic neutron scattering. We hope they will serve as a demo for genuine polarized analysis in future experiments.

\section{Supplemental Numerical Results}

\subsection{Additional results for \texorpdfstring{$\Delta=0.25$}{Delta=0.25}}

\begin{figure}
	\includegraphics[width=\columnwidth]{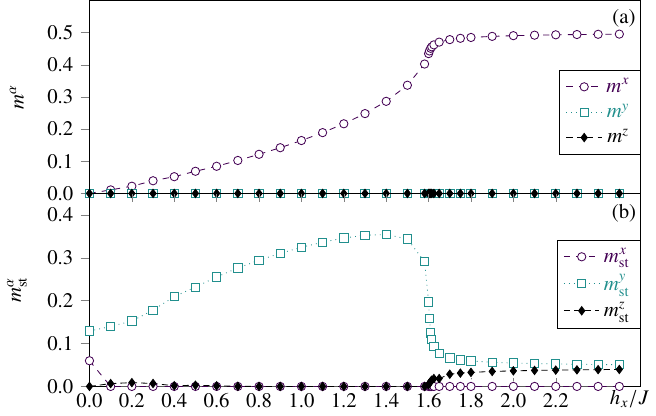}
	\caption{\label{fig:magnetizationComponents}Magnetization components. (a) All magnetization components $m^\alpha$ and (b) staggered magnetization components $m^\alpha_\mathrm{st}$, where $\alpha\in\{ x,y,z\}$, for $\Delta=0.25$.
	}
\end{figure}

Fig.~\ref{fig:magnetizationComponents} shows all computed magnetization and staggered magnetization components. Similarly to Ref.~\cite{PhysRevB.94.085136} we observe some small but finite staggered magnetization both at $h_x=0$ and $h_x>h_\mathrm{c}$, which may be attributed to finite-size effects with open boundary conditions.

\begin{figure}
    \includegraphics[width=\columnwidth]{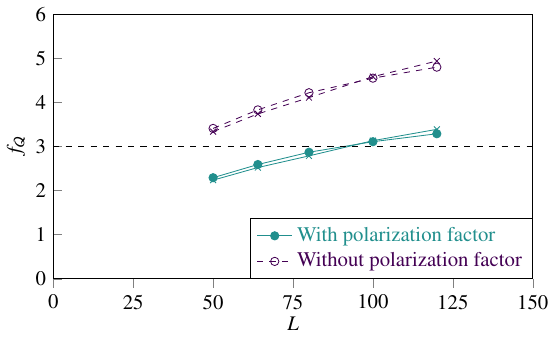}
    \caption{\label{fig:QFIsize}Finite-size scaling of $f_\mathcal{Q}$. QFI density calculated at zero field ($h_x=0$) as a function of system size $L$. For these calculations, the broadening and number of kept states were changed according to $\eta\propto 1/L$ and $m\propto L$. The QFI continues to grow beyond $L=100$.}
\end{figure}

Fig.~\ref{fig:QFIsize} shows zero-field QFI density as a function of system size. We note that $f_\mathcal{Q}$ is expected to diverge with system size at quantum critical points \cite{Hauke2016}. Here we find that the QFI density continues to grow beyond $L=100$ sites, which is the size used for the majority of our results. To achieve finite-size scaling of the dynamical correlation functions used to calculate $f_\mathcal{Q}$, we scaled $\eta \propto 1/L$ with $\eta=0.1$ for $L=100$ as described in Refs.~\cite{PhysRevB.66.045114,PhysRevE.94.053308}. We also modified the number of kept states according to $m\propto L$, with $m=1000$ for $L=100$.

\begin{figure*}
	\includegraphics[width=\textwidth]{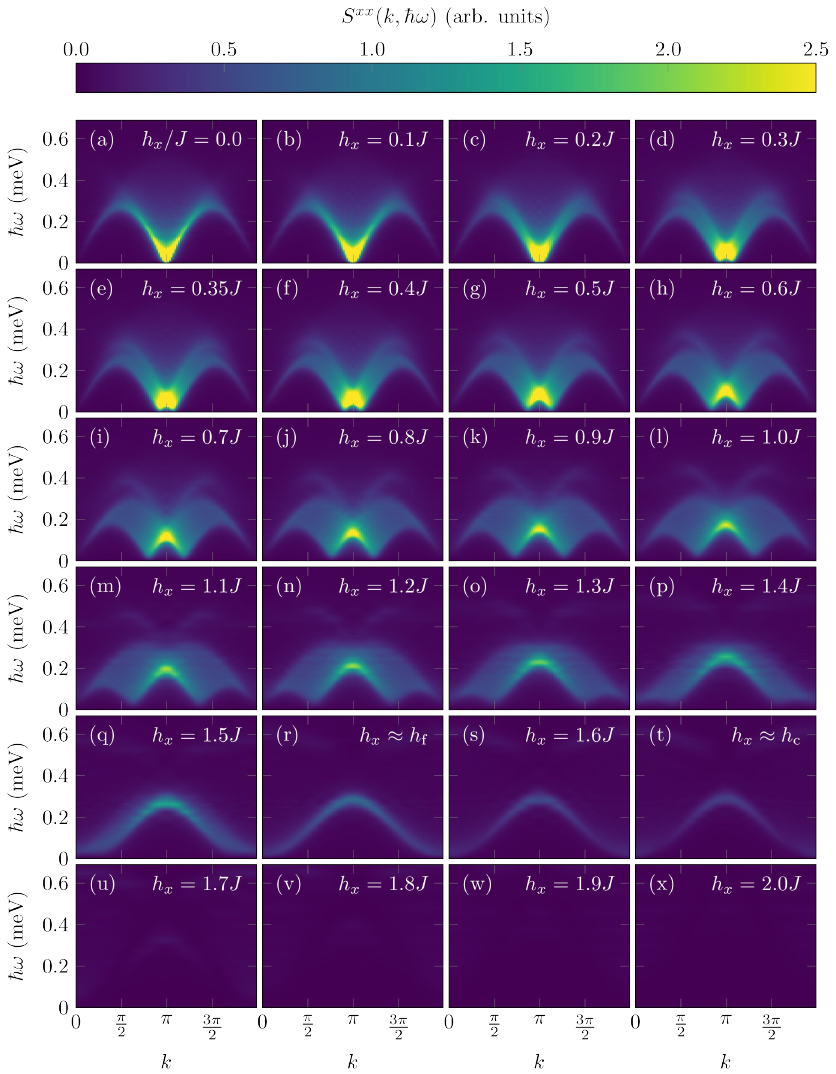}
	\caption{\label{fig:spectra:XXZ:Sxx}$S^{xx}(k,\hbar\omega)$ as a function of the transverse field $h_x$ for the XXZ model with $J=0.23$ meV, $\Delta=0.25$, and elastic features removed.}
\end{figure*}

\begin{figure*}
	\includegraphics[width=\textwidth]{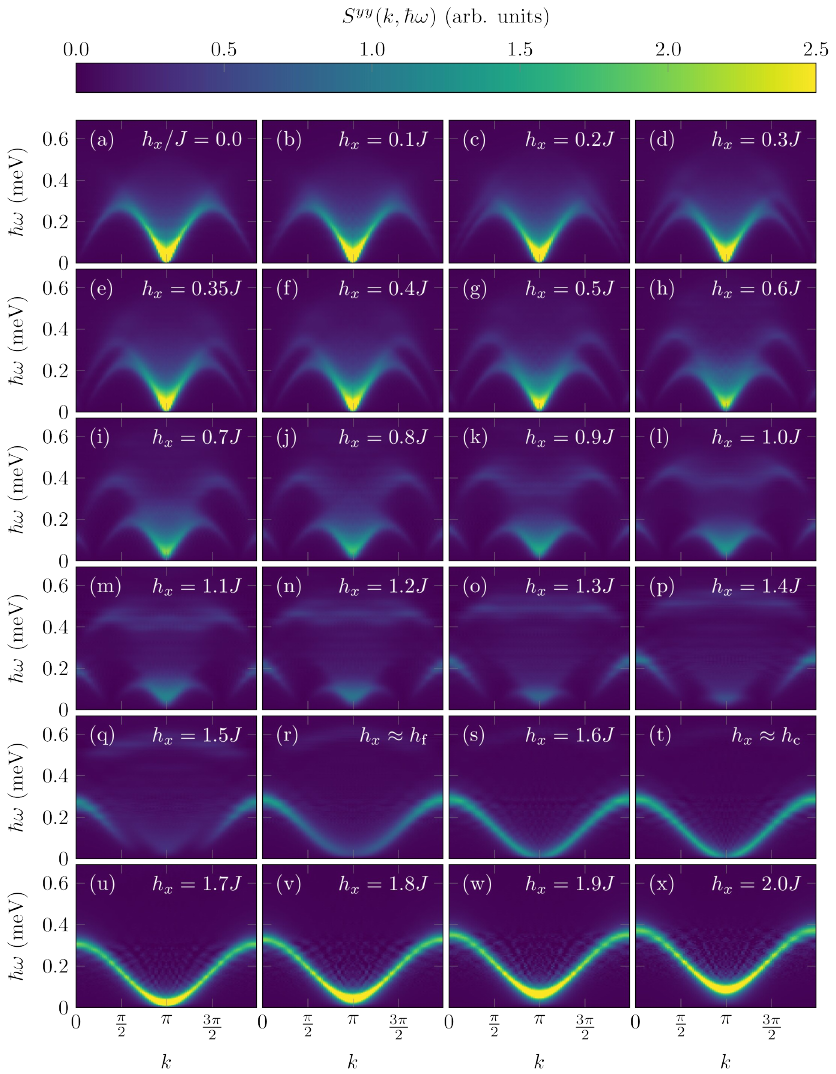}
	\caption{\label{fig:spectra:XXZ:Syy}$S^{yy}(k,\hbar\omega)$ as a function of the transverse field $h_x$ for the XXZ model with $J=0.23$ meV, $\Delta=0.25$, and elastic features removed.}
\end{figure*}

\begin{figure*}
	\includegraphics[width=\textwidth]{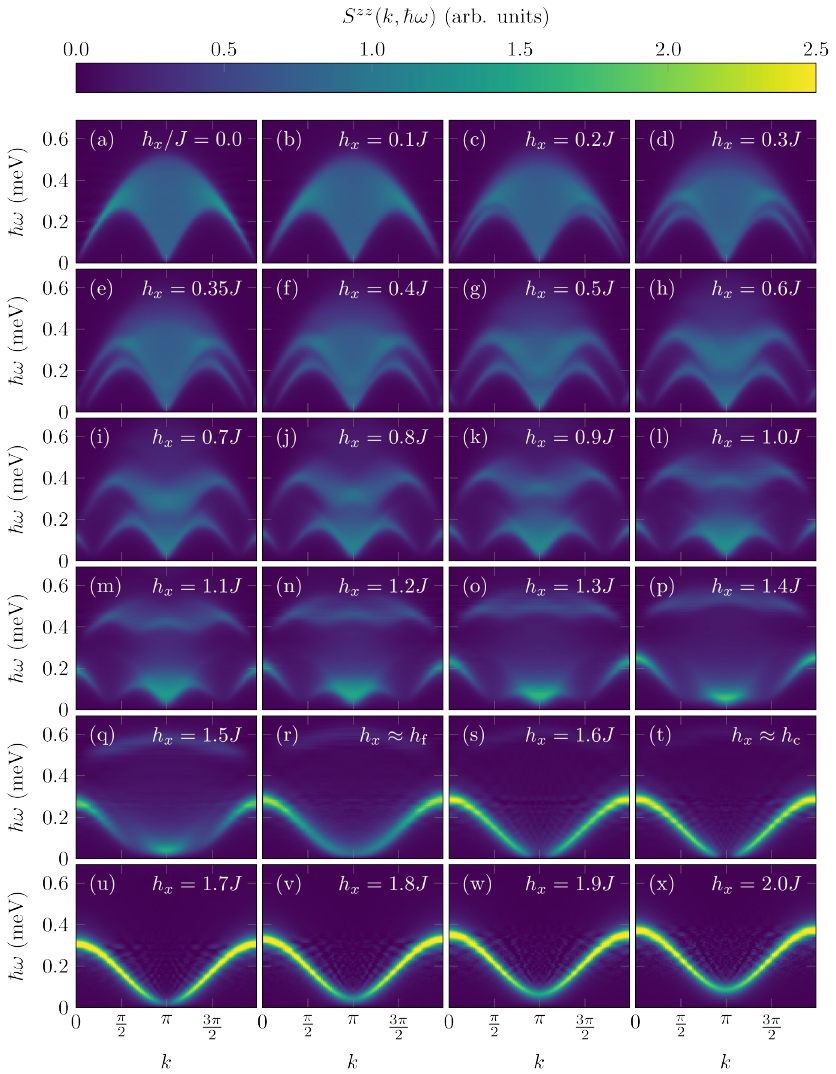}
	\caption{\label{fig:spectra:XXZ:Szz}$S^{zz}(k,\hbar\omega)$ as a function of the transverse field $h_x$ for the XXZ model with $J=0.23$ meV, $\Delta=0.25$, and elastic features removed.}
\end{figure*}

\begin{figure*}
	\includegraphics[width=\textwidth]{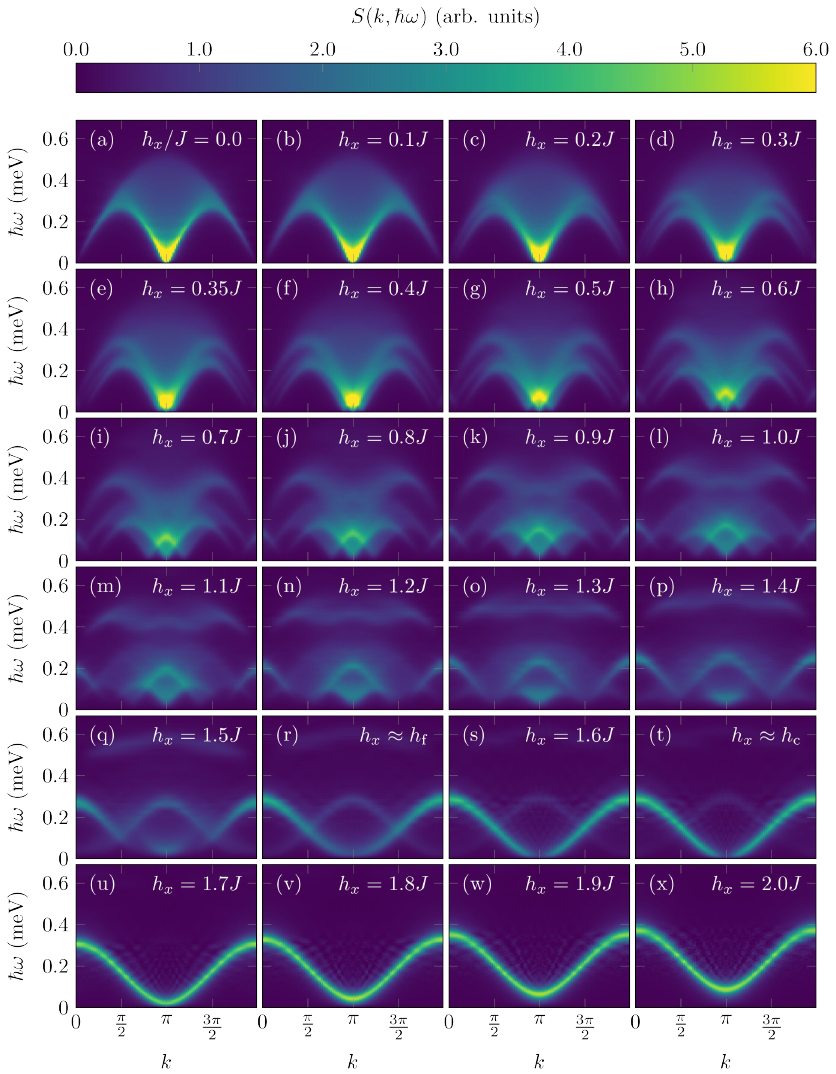}
	\caption{\label{fig:spectra:XXZ:Sqw}Total dynamical spin structure factor. $S(k,\hbar\omega) = \sum_\alpha S^{\alpha\alpha}(k,\hbar\omega)$ as a function of the transverse field $h_x$ for the XXZ model with $J=0.23$ meV, $\Delta=0.25$, and elastic features removed.}
\end{figure*}

We also present the constituent $S^{\alpha\alpha}(k,\hbar\omega)$, $\alpha\in \{x,y,z\}$ components of the spectra shown in the main text, and used for the computation of the QFI field dependence. Figs.~\ref{fig:spectra:XXZ:Sxx}, \ref{fig:spectra:XXZ:Syy} and \ref{fig:spectra:XXZ:Szz} show the evolution of $S^{xx}(k,\hbar\omega)$, $S^{yy}(k,\hbar\omega)$, and $S^{zz}(k,\hbar\omega)$, respectively. For completeness, the full theoretical structure factors $S(k,\hbar\omega)=\sum_\alpha S^{\alpha\alpha}(k,\hbar\omega)$ are shown in Fig.~\ref{fig:spectra:XXZ:Sqw}.

\subsection{Results for other values of \texorpdfstring{$\Delta$}{Delta}}

\begin{figure}
	\includegraphics[width=\columnwidth]{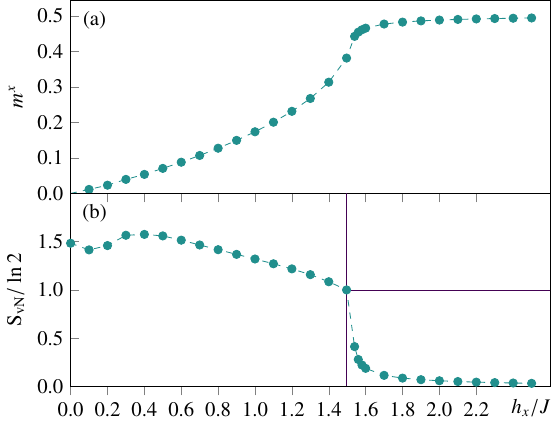}	
	\caption{\label{fig:Delta012}Results for $J=0.23$ meV and $\Delta=0.12$. (a) Magnetization and (b) entanglement entropy as functions of field. The vertical line indicates the factoring field, and the horizontal line is drawn at $\ln 2$.
	}
\end{figure}

\begin{figure}
	\includegraphics[width=\columnwidth]{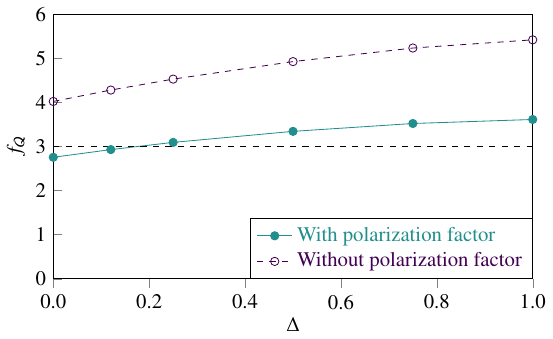}
	\caption{\label{fig:QFIvsDelta}Zero-field QFI. Quantum Fisher information, $f_\mathcal{Q}$, from DMRG simulation as a function of XXZ anisotropy $\Delta$ at $h_x=0$. All other parameters (including $L=100$, $\eta$, $\Delta \omega$) were kept constant, resulting in a suppression of $f_\mathcal{Q}$ at low $\Delta$ due to a decreased bandwidth of excitations. Results are shown with and without applying the polarization factor from the Cs$_2$CoCl$_4$ experiment. Above the dashed horizontal line, QFI indicates more than bipartite entanglement.}
\end{figure}

\begin{figure}
	\includegraphics[width=\columnwidth]{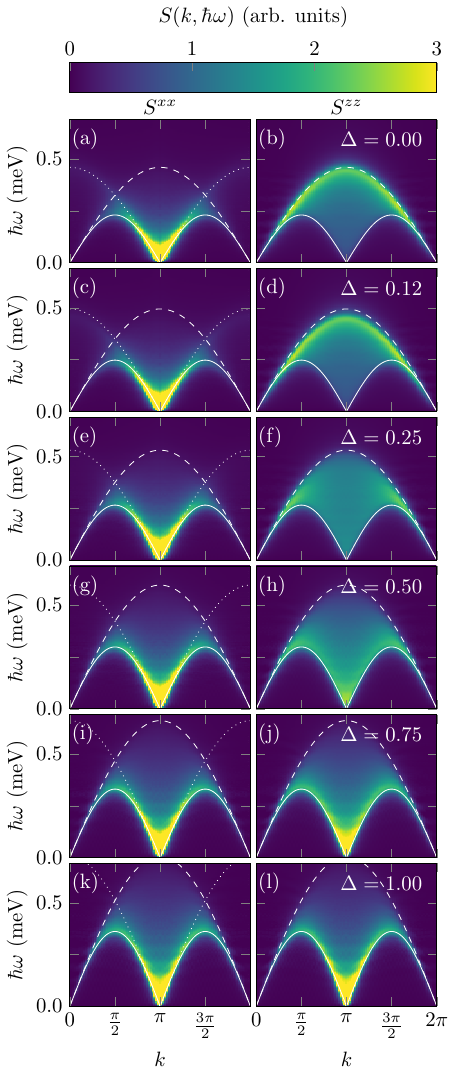}
	\caption{\label{fig:h000:twospinoncontinua}Zero-field spectra. Evolution as function of anisotropy $\Delta$ of $S^{xx}(k,\hbar\omega) = S^{yy}(k,\hbar\omega)$ (left column) and $S^{zz}(k,\hbar\omega)$ (right column) at zero-field for the XXZ chain with $J=0.23$ meV. The solid and dashed yellow white lines show the lower ($\epsilon_1(k)$) and upper bounds ($\epsilon_1(k)$) of the first continuum, respectively. The dashed lines show the upper bound of the second continuum, $\tilde{\epsilon}_2(k)$.
	}
\end{figure}

While our focus has been on $\Delta=0.25$ due to its relevance to Cs$_2$CoCl$_4$, we have also used DMRG to explore other anisotropy values $0\le \Delta \le 1$ to check that our results do not qualitatively depend on fine tuning. First, we provide magnetization and entanglement entropy results for $\Delta=0.12$ in Fig.~\ref{fig:Delta012}. Second, Fig.~\ref{fig:QFIvsDelta} shows the zero-field QFI as a function of $\Delta$. While we see some suppression of $f_\mathcal{Q}$ in the XY-limit, this may be due to a the decreased continuum bandwidth as $\Delta$ is decreased. Since the CV calculation uses a fixed frequency spacing $\Delta \omega$, this bandwidth has to be contained in a lower number of discrete frequencies, which could increase the impact of subtracting elastic peaks on intense scattering at low $\hbar\omega>0$. Importantly, however, we find that it is possible to observe at least bipartite entanglement in the XXZ chain throughout the easy-plane anisotropy regime.

Finally, we show the zero-field DSF components in Fig.~\ref{fig:h000:twospinoncontinua} as a function of the anisotropy. Also shown are bounds for the two-spinon continua. At $h_x=0$ and $T=0$, the main contributions to the dynamical spin structure factor are expected to come from the two-spinon continuum \cite{Muller1981,PhysRevLett.95.077201}. For $S^{zz} \left( k,\hbar\omega\right)$ the bounds of the two-spinon continuum are given by \cite{Muller1981,PhysRevB.55.5833}
\begin{align}
\epsilon_1 (k)	&=	\frac{\pi J \sin \left( \gamma\right)}{2\gamma} \sin k,	\\
\epsilon_2 (k)	&=	\frac{\pi J \sin \left( \gamma\right)}{\gamma} \sin \frac{k}{2},
\end{align}
where $\cos \gamma = \Delta$. The continuum is defined as the energy range $\epsilon_1 < \epsilon < \epsilon_2$. For $S^{xx}\left( k,\hbar\omega\right)$, there are two contributing continua. The first is bounded by $\epsilon_1$, $\epsilon_2$ above. The second is obtained by replacing $k\rightarrow \pi - k$ above, and thus has the bounds
\begin{align}
\tilde\epsilon_1 (k)	&=	\epsilon_1 \left( \pi - k\right) = \epsilon_1,	\\
\tilde\epsilon_2 (k)	&=	\epsilon_2 \left( \pi - k\right) = \frac{\pi J \sin \left( \gamma\right)}{\gamma} \cos \frac{k}{2}.
\end{align}

\section{Reproducing the numerical results.}

The \textsc{DMRG}++ computer program \cite{Alvarez2009} can be obtained with:

\begin{small}
\begin{verbatim}
git clone https://github.com/g1257/dmrgpp.git
\end{verbatim}
\end{small}
and PsimagLite with:
\begin{small}
\begin{verbatim}
git clone https://github.com/g1257/PsimagLite.git
\end{verbatim}
\end{small}
To compile:
\begin{small}
\begin{verbatim}
cd PsimagLite/lib; perl configure.pl; make
cd ../../dmrgpp/src; perl configure.pl; make
\end{verbatim}
\end{small}
For convenience, in the following we define
\begin{small}
\begin{verbatim}
export PSC=/path/to/dmrgpp/scripts
\end{verbatim}
\end{small}

The documentation can be found at 
\nolinkurl{https://g1257.github.io/dmrgPlusPlus/manual.html} or can be obtained
by doing \verb!cd dmrgpp/doc; make manual.pdf!.

To reproduce Figure 1 of the main text, excited state runs are needed. This can be done with \verb!./dmrg -f 100all.ain! where all inputs are provided in \verb!Inputs.tar.gz!. The energies and expectation values are contained in the .cout output file. For the other figures, ground state runs are needed. First run \verb!./dmrg -f 100gs.ain!. After the ground state run has finished, the entanglement entropy plotted in Fig. 4(a) can be extracted from the .cout file. The correlation functions used in the $\tau_2$ calculations can be obtained with
\begin{small}
\begin{verbatim}
./observe -f 100gs.ain "<X0|sz;sz|X0>"  > SzSz.txt
\end{verbatim}
\end{small}
for $\langle S_i^z S_j^z \rangle$ (and similarly for $\langle S_i^x S_j^x \rangle$, $\langle S_i^y S_j^y \rangle$). To symmetrize the output,
\begin{scriptsize}
\begin{verbatim}
perl ${PSC}/staticCorrelations.pl SzSz.txt sz > SzSz_symm.txt
\end{verbatim}
\end{scriptsize}
etc., after which further processing can be done.

The batches and inputs for all frequency runs can be generated with
\begin{tiny}
\begin{verbatim}
perl -I${PSC} ${PSC}/manyOmegas.pl inputDollarized.ado XXZ_Sqw.pbs test
\end{verbatim}
\end{tiny}
that can be launched by replacing \verb!test! with \verb!submit!. The $S^{xx}(k,\omega)$, $S^{yyx}(k,\omega)$, and $S^{zz}(k,\omega)$ are run separately. When the frequency runs for one component has finished, the post-processing to obtain the corresponding \verb!.pgfplots! 
file is
\begin{scriptsize}
\begin{verbatim}
perl -I${PSC} ${PSC}/procOmegas.pl -f InputDollarized.ado -p
\end{verbatim}
\end{scriptsize}
after which further processing can be done.

\section{Mathematical Notes}

This appendix states and proves mathematical statements made in the main paper.
We consider a system of $N$ spins, each with $s=1/2$, 
on a bipartite lattice, and a tensor product of $N$ Hilbert spaces, each of dimension two.
We consider
\begin{equation}
H = \sum_{i,j} J_{i,j} \left(S^{x}_i S^x_j + S^{y}_i S^y_j + \Delta S^{z}_i S^z_j\right) + h_x\sum_i S^x_i,
\label{eq:Hjij}
\end{equation}
which is slightly more general than Eq.~(1) of the main text, and where $J_{i,j}$ is real,  \emph{only} connects
sites of different sublattices, and where $J_{i,i}=0$.

For $h_x=0$ and when $J_{i,j}=J$ for nearest neighbors and zero otherwise, the 1D ground-state energy per site in the thermodynamic limit can be obtained exactly from the 
Bethe ansatz \cite{PhysRev.147.303,PhysRev.150.327},
\begin{align}
\frac{E_0}{L}	&=	\left\{ \begin{array}{rl}
\frac{\Delta}{4} & \text{for } \Delta \le -1,\\
\frac{\Delta}{4} - \frac{1}{2} \left( 1-\Delta^2\right) \int_{-\infty}^\infty \frac{dx}{\cosh \left( \pi x\right) 
	\left[ \cosh \left( 2x \arccos \Delta \right) -\Delta \right]} & \text{for } \Delta > -1,	\\
\frac{1}{4} - \log 2	& \text{for } \Delta = 1.
\end{array}\right.
\end{align}
For $\Delta = 0.25$ we obtain $E_0/L=-0.3451797$.

In the thermodynamic limit, there exists  a critical field $h_\mathrm{c} > 0$, such that $H$ has a doubly-degenerate 
ground state for $|\Delta|<1$ and $0<h_x<h_\mathrm{c}$; a staggered $y-$magnetization appears in this parameter range, and disappears 
for $h\ge h_\mathrm{c}$; the $x$-magnetization increases with increasing $h_x$, and saturates at $h_x = h_\mathrm{c}$.

For finite $N$, this is not so:
For $h_x=0$ and $\Delta\ge -1$, $H$ has a ground state that is non-degenerate \cite{doi:10.1063/1.1724276, Affleck1986}.
By continuity, there must exist $h_x^0 > 0$ such that $\forall\,\,0\le h_x < h_x^0,$ the ground state of $H$ is non-degenerate.
Therefore, to approach the thermodynamic limit from finite systems, we measure $\langle 0|S^y_r|1\rangle$ as a proxy for
the $y-$magnetization, check that the first gap $E_1 - E_0$ quickly decreases to zero with increasing $N$, and calculate the second gap $E_2-E_0$
as a proxy for ``the gap'' in the thermodynamic limit. In what follows, we assume finite dimensional Hilbert
spaces over the complex numbers, unless otherwise noted.\\

\mytheorem{} Let $H$ and $T$ be commuting matrices.
Let $|n\rangle$ be a \emph{non-degenerate} eigenvector of $H$ with
eigenvalue $E_n$. Then $|n\rangle$ is also an eigenvector of $T$.

\myproof\, Just note that $T|n\rangle$ is an eigenvector of $H$ with non-degenerate eigenvalue $E_n$.\\

Let $T_r$ be the translation operator of $r$ sites.
Note that
$T^{-1}_r = T_{-r}$, because translating back undoes the translating forward, 
$T_{r+p} = T_r T_p$, 
and the eigenvalues of $T_r$ have norm 1.

\mytheorem[lemma] $T^{-1}_{r} S_r T_{r} = S_0$

\myproof\,Let $S_r$ be a local observable, so that $S_r = I \otimes \cdots S \otimes I \otimes \cdots I$
where S appears \emph{only} at location $r.$
Consider just three sites, and consider $r=1.$
\begin{align}
T^{-1}_{1} S_1 T_{1} (|a\rangle \otimes |b\rangle \otimes |c\rangle) &= T^{-1}_{1} S_1 (|c\rangle \otimes |a\rangle \otimes |b\rangle) =\nonumber\\
T^{-1}_{1} (|c\rangle \otimes S|a\rangle \otimes |b\rangle) &=
 S|a\rangle \otimes |b\rangle \otimes |c\rangle = S_0 (|a\rangle \otimes |b\rangle \otimes |c\rangle)\nonumber
\end{align}

\mytheorem{} If $H$ is translation invariant, and $|n\rangle$
an eigenvector of $H$ with \emph{non-degenerate} eigenvalue $E_n$, 
then $\langle n| S_r | n\rangle = \langle n|S_0|n\rangle$ for all sites $r$,
where $S_r = I \otimes \cdots S \otimes I \otimes \cdots I$, with $S$ any operator acting \emph{only} on site $r.$

\myproof\, Because $|n\rangle$ is non-degenerate, then $|n\rangle$ is an eigenstate of $T_r$
with eigenvalue $\lambda$, where $\lambda$ is complex of norm 1.
\begin{equation}
\langle n| S_r | n\rangle = \langle n| \lambda^* S_r \lambda |n\rangle = \langle n|T_{-r} S_r T_r |n\rangle = \langle n|S_0|n\rangle
\end{equation}
where the \reflemma\, was used in the last step.

Remark: Note that if ket and bra are different states, like $\langle n| S_r | m\rangle$ with $n \neq m$, this
proposition no
longer holds, because $|n\rangle$ and $|m\rangle$ might have different $T_r$ eigenvalues.\\

\mytheorem{} If $H$  given by Eq.~\eqref{eq:Hjij} is
translation invariant, then  $\langle n|S^y_r|n\rangle = 0$ for all
\emph{non-degenerate} energy eigenstates $n$, and all sites $r$.

\myproof\, If $h_x = 0$ and $n$ is non-degenerate,
then $|n\rangle$ is also an eigenvector of $S^z\equiv\sum_r S^z_r$, and 
$\langle n|S^y_r|n\rangle = 0$ follows.
Translation invariance is not necessary in this case.

We now consider $h_x \neq 0$.
For any fixed site $r,$ it is true that $\langle n|[H, S^z_r]|n\rangle = 0$.
Note that the $zz$ term of $H$ does not appear in this commutator, so no $\Delta$ appears there.
\begin{align}
0 = \langle n|[H, S^z_r]|n\rangle = \sum_{p} J_{r,p} \langle n|S^x_{p} S^y_r - S^x_r S^y_{p}|n\rangle - hx \langle n|S^y_r|n\rangle
\nonumber
\end{align}

Because $J_{r,r} = 0$ (at equal sites) then $p \neq r,$ and thus $S^x_{p} S^y_r = S^y_r S^x_{p}$.
Now summing over $r$,
\begin{align}
0 = \sum_{r,p} J_{r,p} \langle n|S^x_{p} S^y_r - S^x_r S^y_{p}|n\rangle 
- h_x \sum_r \langle n|S^y_r|n\rangle
\label{eq:commutator}
\end{align}
The first term of Eq.~\eqref{eq:commutator} is zero, because $J_{r,p} = J_{p,r}$, so
that its second term is zero also, which implies (because $|h_x| > 0$) that
$\sum_r \langle n|S^y_r|n\rangle = 0$.
Finally $\langle n|S^y_r|n\rangle$  does \emph{not} depend on $r$, because 
$H$ is translation invariant, $n$ is non-degenerate, and $S^y_r$ is local, and then
$\langle n|S^y_r|n\rangle = 0.$\\

\mytheorem{}
If the ground state of Eq.~\eqref{eq:Hjij} is non-degenerate, the Hamiltonian is translation invariant, and
there is a non-zero ground state $x-$magnetization, then an elastic peak ($\omega=0$) is present in $S^{xx}(k=0, \omega=0)$.

\myproof\,Let $|n\rangle$ be the eigenvectors of H with eigenvalues $E_n$, and we assume $|h_x| > 0$.
Then the
weight $W^{xx}$ of $S^{xx}(k, \omega = 0)$ is given by
\begin{align}
W^{xx}(k) = \sum_{n; E_n = E_0} \sum_{r, p} \textrm{e}^{ik(r-p)} \langle 0|S^x_r|n\rangle\langle n|S^x_{p}|0\rangle
\end{align}

Assuming the only state with energy (almost) the same as $E_0$ is the first excited $|1\rangle,$ then
\begin{align}
W^{xx}(k) = \sum_{r, p} \textrm{e}^{ik(r-p)} [\langle 0|S^x_r|0\rangle\langle 0|S^x_{p}|0\rangle + 
\langle 0|S^x_r|1\rangle\langle 1|S^x_{p}|0\rangle]\nonumber
\end{align}
Let
\begin{align}
g_n(k) \equiv \sum_{r, p} \textrm{e}^{ik(r-p)}\langle 0|S^x_r|n\rangle\langle n|S^x_{p}|0\rangle
\end{align}
Then $W^{xx}(k) =  g_0(k) + g_1(k)$.
Assuming PBC and that $|0\rangle$ is non-degenerate, then all points $r$ are equivalent 
so that $ \langle 0|S^x_r|0 \rangle =  \langle 0|S^x_0|0\rangle$ for all sites $r.$
Then
$g_0(k) = |\langle 0|S^x_0|0\rangle|^2  N  \delta(k, 0)$.
(It is \emph{incorrect} to assume that $\langle 0|S^x_r|1\rangle$
 doesn't depend on $r,$ because $|0\rangle$ and $|1\rangle$ are \emph{different} states.)

On the other hand,
\begin{align}
g_1(k=0) =  \sum_{r, p}\langle 0|S^x_r|1\rangle \langle 1|S^x_{p}|0\rangle = 
\left|\sum_r \langle 0|S^x_r|1\rangle \right|^2 \ge 0
\end{align}
So that $W^{xx}(k = 0) \ge  N|\langle 0|S^x_0|0\rangle |^2 > 0$
because the $x-$magnetization $|\langle 0|S^x_0|0\rangle | > 0$ \emph{strictly}.
This proves that there is an elastic peak ($\omega=0$) for $S^{xx}$ at $k = 0.$\\

\mytheorem{}
If the ground state of Eq.~\eqref{eq:Hjij} is non-degenerate, the Hamiltonian is translation invariant, and
$|\langle 0|S^y_r|1\rangle| > 0,$ then an elastic peak ($\omega=0$) is present in
 $S^{yy}(k, \omega=0)$ for one or more $k$ values.

\myproof\,We repeat the above, and using that (because of symmetry) $\langle 0|S^y_r|0\rangle = 0$ for all $r,$ we get
that the peak of $S^{yy}$ at $\omega=0$ is
\begin{align}
W^{yy}(k) =  \sum_{r, p}\textrm{e}^{ik(r-p)} \langle 0|S^y_r|1\rangle\langle 1|S^y_{p}|0\rangle
\end{align}
(It is \emph{incorrect} to assume that $\langle 0|S^y_r|1\rangle$
doesn't depend on $r,$ because $|0\rangle$ and $|1\rangle$  are \emph{different} states.)

We can sum over $k$ though, and then  $\sum_k \textrm{e}^{ik(r-p)}= \delta(r, p),$ and we get
\begin{align}
\sum_k S^{yy}(k, \omega = 0) =  \sum_{r} \langle 0|S^y_r|1\rangle\langle 1|S^y_r|0\rangle =  \sum_{r} |\langle 0|S^y_r|1\rangle|^2
\end{align}
which is \emph{strictly} greater than zero, because it is the $y-squared$ magnetization.

Therefore there \emph{must} be an elastic peak ($\omega=0$) also for $S^{yy},$ for at least one k.
(Using other arguments, we know this $k$ is $\pi$.)

\mytheorem{} (M\"uller and Shrock, 1985) If $\Delta> 0$ and $h=J\sqrt{2(1+\Delta)}$, then the ground state of Eq.~\eqref{eq:Hjij} is given by
a product state, the \emph{classical point}, with energy $E/J=-J(2+\Delta)/4$; this ground state is at least doubly degenerate if $\theta\neq0$,
where $J\cos\theta = h/(1+\Delta)$. Moreover, 
\begin{align}
S_{xx}(k,\omega) &= S_{zz}(k-\pi, \omega)\sin^2\kern-2pt\theta + \pi^2\cos^2\kern-2pt\theta\,\delta(\omega)\,\delta(k),\nonumber\\
S_{yy}(k,\omega) &= S_{zz}(k, \omega)\cos^2\kern-2pt\theta + \pi^2\sin^2\kern-2pt\theta\,\delta(\omega)\,\delta(k-\pi).\label{eq:mullerspectral}
\end{align}

\myproof\, See \cite{PhysRevB.32.5845}.

\mytheorem{}
The Hamiltonian
\begin{equation}
H = \sum_{i,j} J_{i,j} \left(S^{x}_i S^x_j + \Delta S^{y}_i S^y_j +  S^{z}_i S^z_j\right) + h_x\sum_i S^x_i
\label{eq:HjijEquiv}
\end{equation}
is equivalent to Eq.~\eqref{eq:Hjij}, so that we can swap $z$ and $y$ \emph{and} add a tilde to $S$ everywhere in Eq.~\eqref{eq:mullerspectral}.

\myproof{}
Apply the canonical transformation $S^{y}=\tilde{S}^z$ and $S^{z}=-\tilde{S}^y$ to Eq.~\eqref{eq:Hjij} to obtain Eq.~\eqref{eq:HjijEquiv}